\documentclass[PAPER,  atlasdraft=false, cernpreprint=true, coverpage=false, texlive=2023, UKenglish, texmf, orcidlogo]{atlasdoc}
 
\usepackage{atlaspackage}
\usepackage{atlasbiblatex}
 
\usepackage{atlasphysics}

\graphicspath{{logos/}{figures/}}
 
\usepackage{ANA-HDBS-2019-13-PAPER-defs}

\AtlasTitle{Search for dark photons from Higgs boson decays via $ZH$ production with a photon plus missing transverse momentum signature from $pp$ collisions at $\sqrt{s}$ = 13 TeV with the ATLAS detector}
 
\AtlasVersion{3.0}
 
\AtlasAbstract{
This paper describes a search for dark photons ($\gamma_d$) in proton-proton collisions at $\sqrt{s}=13$~TeV at the Large Hadron Collider (LHC). The dark photons are searched for in the decay of Higgs bosons ($H \to \gamma\gamma_d$) produced through the $ZH$ production mode. The transverse mass of the system, made of the photon and the missing transverse momentum from the non-interacting $\gamma_d$, presents a distinctive signature as it peaks near the Higgs boson mass. The results presented use the total Run-2 integrated luminosity of 139~fb$^{-1}$ recorded by the ATLAS detector at the LHC. The dominant reducible background processes are estimated using data-driven techniques. A Boosted Decision Tree technique is adopted to enhance the sensitivity of the search. As no excess is observed with respect to the Standard Model prediction, an observed (expected) upper limit on the branching ratio BR$(H\to \gamma\gamma_d)$ of 2.28$\%$ (2.82$^{+1.33}_{-0.84}\%$) is set at 95$\%$~CL for massless $\gamma_d$. For massive dark photons up to 40~GeV, the observed (expected) upper limits on  BR$(H\to \gamma\gamma_d)$ at 95\%~confidence level is found within the [2.19,2.52]$\%$ ([2.71,3.11]$\%$) range.
}

\AtlasRefCode{HDBS-2019-13-PAPER}
 
\PreprintIdNumber{CERN-EP-2022-237}

\AtlasJournal{JHEP}
\AtlasJournalRef{JHEP 07 (2023) 133}
\AtlasDOI{10.1007/JHEP07(2023)133}

\hypersetup{pdftitle={ATLAS document},pdfauthor={The ATLAS Collaboration}}
 
\begin{document}
 
\maketitle
 
\tableofcontents
\newpage

\section{Introduction}
\label{sec:intro}

 
There is strong astrophysical evidence suggesting the existence of dark matter (DM) with a density about five times higher than ordinary baryonic matter \cite{Clowe_2006}. However, its fundamental nature is unknown and it is plausible that it is a component of a larger ``dark sector'', which couples weakly to the Standard Model (SM) and possesses a rich internal structure and interactions. Existing models propose dark sectors that contain few or many particles, some providing ideal candidates for DM. The interaction of these dark states can be Yukawa-like or mediated by dark gauge bosons or both.
 
The dark and visible sectors may interact through a portal offering a potential experimental signature. The form of this portal can be classified according to the type and dimension of its operators. The best motivated and most studied cases contain relevant operators taking different forms, depending on the spin of the mediator: vector (spin 1), neutrino (spin 1/2), Higgs boson (scalar) and axion (pseudo-scalar) \cite{Fabbrichesi_2021}.
 
The vector portal considered in this paper is the one where the interaction results from the kinetic mixing between one dark and one visible Abelian gauge boson. The visible photon is taken to be the boson of the $U(1)$ gauge group of electromagnetism---or the hyper-charge in the regime above the electroweak symmetry breaking scale---and the dark photon ($\gamma_d$) is identified as the boson of an extra $U(1)_D$ gauge group of the dark sector. This mixing is always possible because the field strengths of two Abelian gauge fields can be multiplied together to give a dimension four operator \cite{Fayet:1990}. The existence of such an operator means that the two gauge bosons mix as they propagate \cite{Georgi:1983}. This kinetic mixing provides the portal linking the dark and visible sectors and makes the experimental detection  of the dark photon possible.
 
Massless and massive dark photons, whose theoretical frameworks as well as experimental signatures are quite distinct, give rise to dark sectors with different phenomenological and experimental features. The massive dark photon has received so far most of the attention because it couples directly to the SM currents and is more readily accessible in the experimental searches \cite{Raggi:2015,Deliyergiyev:2015,DarkSector2022,Curciarello:2016,Beacham:2020}. The massless dark photon arises from a sound theoretical framework \cite{Fabbrichesi_2021}. It does not couple directly to any of the SM currents and interacts instead with ordinary matter only through operators of dimension higher than four. It provides, with respect to the massive case, a comparably rich, if perhaps more challenging, experimental target.
 
Looking beyond particle physics and towards cosmology, dark photons may solve the small-scale structure formation problems~\cite{Fan:2013}. In astroparticle physics, dark photons may induce the Sommerfeld enhancement of the DM annihilation cross-section needed to explain the PAMELA-Fermi-AMS2 positron anomaly \cite{Arkani:2009}. They may also assist light DM annihilation to reach the phenomenologically required magnitude, and make asymmetric DM scenarios phenomenologically viable  \cite{Gabrielli:2014oya, Biswas:2016jsh}.
 
This analysis searches for dark photons predicted by a new model generating exponentially spread SM Yukawa couplings from unbroken $U(1)_D$ quantum numbers in the dark sector \cite{Gabrielli_Raidal:2014}. In this approach, non-perturbative flavour- and chiral-symmetry breaking is transferred from the dark to the visible sector via heavy scalar messenger fields that might produce new physics signals at the Large Hadron Collider (LHC). For massless dark photons, the $U(1)_D$ kinetic mixing with $U(1)$ can be tuned away on shell, in agreement with all existing constraints \cite{Kelly:2019,Jaeckel:2013}, while off-shell contributions give rise to higher-dimensional contact operators strongly suppressed by the scale of the heavy messenger mass.
 
A potential discovery process for dark photons proceeding via Higgs-boson production at the LHC is presented in this paper. Thanks to the non-decoupling properties of the Higgs boson, a branching ratio of $H \to \gamma \gamma_d$ with values up to a few percent are possible for a massless dark photon as well as for heavy dark-sector scenarios~\cite{Gabrielli:2014oya,Gabrielli_Raidal:2014,Biswas:2016jsh}. The corresponding signature consists, for a Higgs boson with a mass $m_H = 125$~\GeV{}, of a photon with an energy $E_{\gamma} = m_{H}/2 $ in the Higgs centre-of-mass frame and a similar amount of missing transverse momentum (\MET) which originates from the escaping $\gamma_d$~\cite{Biswas:2016jsh}.
 
Moreover, in the unbroken $U(1)_D$ scenarios, the two-body decay $H \to \gamma \gamma_d$ can be enhanced despite existing theoretical constraints~\cite{Gabrielli:2014oya}, providing a very distinctive  signature of a single photon plus missing transverse momentum  at the Higgs boson mass resonance.  If such a signature is discovered at the LHC, CP invariance will imply the spin-1 nature of the missing particle, excluding axions or other ultra light scalar particles.
 
The photon plus {\MET} signature has been extensively studied by the LHC experiments \cite{EXOT-2018-63,CMS-EXO-16-039,susy-2019-14}. In the particular case of massless $\gamma_d$, searches in Higgs boson decays were performed at the LHC in $pp$ collisions at a centre-of-mass energy of 13~\TeV{}. The CMS experiment has probed this decay channel using Higgs boson events produced in association with a $Z$ boson $ZH(Z\to \dil$) with an integrated luminosity of 137~fb$^{-1}$ \cite{CMS-EXO-19-007}, or via vector-boson fusion (VBF) production \cite{CMS-EXO-20-005} (with 130~fb$^{-1}$) setting an observed (expected) upper limit on the BR($H \to \gamma \gamma_d$) at the 95\%~confidence level~(CL) of 4.6\% (3.6\%) and 3.5\% (2.8\%) respectively. ATLAS has set an  observed (expected) limit on the $H \to \gamma \gamma_d$ branching ratio, using the VBF production mode with an integrated luminosity of 139~fb$^{-1}$, to 1.8\% (1.7\%) at the 95\%~CL \cite{EXOT-2021-17}.
 
\begin{figure}[ht]
\centering
\includegraphics[width=1.\textwidth]{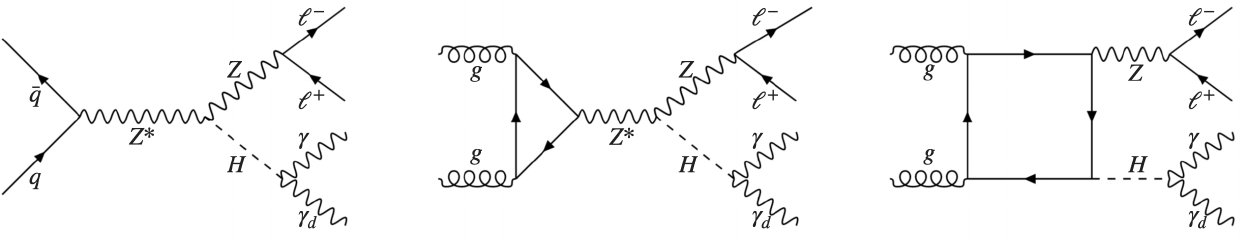}
\caption{Feynman diagrams for H$\rightarrow\gamma\gamma_{d}$ in $q\bar{q} \to ZH$ and $gg \to ZH$ production modes.}
\label{fig:zhyyd}
\end{figure}

This analysis is based on the $ZH$ production mode where $Z \to \dil$ ($\ell=e, \mu$) and $H \to \gamma \gamma_d$, which proceeds at leading order through the Feynman diagrams shown in Figure~\ref{fig:zhyyd}. The study is performed using a final state consisting of two same-flavour, opposite-charge electrons or muons, an isolated photon and missing transverse momentum. The requirements applied to the photon and the \met, originating from a potential SM Higgs boson decay, are optimised to maximise the signal acceptance. The leptons, on the other hand, are used for triggering on the event and provide a $Z$ boson mass constraint. The transverse mass \mt{} of the $\gamma-$\met system presents a kinematic edge at the Higgs boson mass and is included as a variable of interest in the boosted decision tree (BDT) score that is exploited to search for a dark photon signal. The kinematics of these events allow the search for low-mass~($\neq 0$) $\gamma_d$. Hence, the analysis is  optimized for dark photon searches in the [0-40]~\GeV{} mass range.
 
\section{The ATLAS detector}
\label{sec:detector}

\newcommand{\AtlasCoordFootnote}{
ATLAS uses a right-handed coordinate system with its origin at the nominal interaction point (IP) in the centre of the detector and the \(z\)-axis along the beam pipe. The $x$-axis points from the IP to the centre of the LHC ring, and the $y$-axis points upwards. Cylindrical coordinates $(r,\phi)$ are used in the transverse plane, $\phi$ being the azimuthal angle around the $z$-axis. The pseudorapidity is defined in terms of the polar angle $\theta$ as $\eta = -\ln \tan(\theta/2)$. Angular distance is measured in units of $\Delta R \equiv \sqrt{(\Delta\eta)^{2} + (\Delta\phi)^{2}}$.}
 
The ATLAS detector~\cite{PERF-2007-01} at the LHC covers nearly the
entire solid angle around the collision point\footnote{\AtlasCoordFootnote}. It consists of an inner tracking detector surrounded by a thin superconducting solenoid, electromagnetic and hadron calorimeters as well as a muon spectrometer incorporating three large superconducting air-core toroidal magnets.
 
The inner-detector system (ID) is immersed in a \SI{2}{\tesla} axial magnetic field and provides charged-particle tracking in the range of pseudorapidity $|\eta|<2.5$.  The high-granularity silicon pixel
detector covers the vertex region and typically provides four measurements per track, the first hit normally being in the insertable
B-layer (IBL) installed before Run~2~\cite{ATLAS-TDR-19,PIX-2018-001}. It is followed by the silicon microstrip tracker (SCT), which usually provides eight measurements per track. These silicon detectors are complemented by the transition radiation tracker (TRT), which enables radially extended track reconstruction up to $|\eta| = 2.0$. The TRT also provides electron identification information based on the fraction of hits (typically 30 in total) above a higher energy-deposit threshold corresponding to transition radiation.
 
The calorimeter system covers the pseudorapidity range $|\eta|<4.9$.  Within the region $|\eta|<3.2$, electromagnetic (EM) calorimetry is provided by barrel and endcap high-granularity lead/liquid-argon (LAr) calorimeters, with an additional thin LAr presampler covering $|\eta|<1.8$ to correct for energy loss in material upstream of the calorimeters. Hadron calorimetry is provided by the steel/scintillator-tile calorimeter, segmented into three barrel structures within $|\eta|<1.7$, and two copper/LAr hadron endcap calorimeters. The solid angle coverage is completed with forward copper/LAr and tungsten/LAr calorimeter modules optimised for electromagnetic and hadronic energy measurements respectively.
 
The muon spectrometer (MS) comprises separate trigger and high-precision tracking chambers measuring the deflection of muons in a magnetic field generated by the superconducting air-core toroidal magnets. The field integral of the toroids ranges between \num{2.0} and \SI{6.0}{\tesla\metre} across most of the detector. Three layers of precision chambers, each consisting of layers of monitored drift tubes, covers the region $|\eta|<2.7$, complemented by cathode-strip chambers in the forward region, where the background is highest. The muon trigger system covers the range $|\eta|<2.4$ with resistive-plate chambers in the barrel, and thin-gap chambers in the endcap regions.
 
Interesting events are selected by the first-level trigger system implemented in custom hardware, followed by selections made by algorithms implemented in software in the high-level trigger~\cite{TRIG-2016-01}.  The first-level trigger selects events from the \SI{40}{\MHz} bunch crossings at a rate below \SI{100}{\kHz}, which the high-level trigger further reduces in order to record events to disk at about \SI{1}{\kHz}.  An extensive software suite~\cite{ATL-SOFT-PUB-2021-001} is used in the reconstruction and analysis of real and simulated data, in detector operations and in the trigger and data acquisition systems of the experiment.


\section{Data samples}
\label{sec:data}

The data used in this paper were collected by the ATLAS experiment from the LHC \pp collisions at $\rts=$~13~\TeV{} during stable beam conditions, with all subdetectors operational~\cite{DAPR-2018-01}. The corresponding total integrated luminosity is \integLumi. The data were recorded with high efficiency using unprescaled trigger algorithms based on the presence of single leptons or di-leptons, where electrons and muons are considered as leptons~\cite{TRIG-2018-01,TRIG-2018-05}.
 
The trigger thresholds are based on the transverse momentum \pT{} of the leptons and are determined by the data-taking conditions during the different periods~\cite{TRIG-2019-04}, particularly by the number of multiple \pp interactions in the same or neighbouring bunch crossings, referred to as \pileup. The number of \pileup interactions ranges from about 8 to 70, with an average of 34. Single-lepton triggers with low \pT{} threshold and lepton isolation requirements are combined in a logical OR with higher-threshold triggers without isolation requirements to give maximum efficiency. The di-lepton triggers require two leptons that satisfy loose identification criteria, with symmetric (symmetric or asymmetric) \pT{} thresholds for electrons (muons). The di-lepton trigger complements the single-lepton trigger to recover between 3\% and 3.6\% signal efficiency depending on the dark photon mass points for combined \Zee{} and \Zmm{} final states.


\section{Simulated event samples}
\label{sec:sim}

Monte Carlo (MC) simulated samples are used to model both the signal and the different background processes using the configurations shown in Table~\ref{tab:mcconfig}. All the samples were generated at a centre-of-mass energy of 13 \TeV{}.
The generated events were processed through a simulation~\cite{SOFT-2010-01} of the ATLAS detector geometry and response using \GEANT~\cite{Agostinelli:2002hh}, and through the same reconstruction software as the collected  $pp$ collision data. Corrections were applied to the simulated events so that the particle candidates' selection efficiencies, energy scales and energy resolutions match those determined from data control samples. In addition, appropriate scale factors corresponding to the fired triggers are applied~\cite{TRIG-2018-01,TRIG-2018-05}. The simulated samples are normalised to the total integrated luminosity, using the corresponding cross-sections computed to the highest order available in perturbation theory. The \pileup effects were modelled using events from minimum-bias interactions generated using \PYTHIA{}8.186~\cite{Sjostrand:2007gs} with the A3 set of tuned parameters~\cite{ATL-PHYS-PUB-2016-017}. They were overlaid onto the simulated hard-scatter events according to the luminosity profile of the recorded data (reweighting procedure). For massive $\gamma_d$ signal samples and the $tW\gamma$ background process, the detector response was simulated using a fast parameterized simulation of the ATLAS calorimeters~\cite{Atlfast3}. For the massless signal samples and all remaining background samples, the full \GEANT{} simulation was used. All simulated samples, except those produced with the \SHERPA{}2.2.1~\cite{Gleisberg:2008ta} event generator, used \textsc{EvtGen} 1.2.0~\cite{Lange:2001uf} to model the decays of heavy-flavour hadrons.
 
\begin{table*}[!htbp]
\centering
\caption{
The configurations used for event generation of signal and background processes. $V$ refers to an electroweak boson ($W$ or $Z/\gamma^*$). The matrix element (ME) order refers to the order in the strong coupling constant of the perturbative calculation in the MC event generation. PDF refers to the parton density librairies used with the generator. The tune refers to the underlying-event tune of the parton shower model.}
\label{tab:mcconfig}
\centering
\resizebox{\textwidth}{!}{
\def\arraystretch{1.5}
\newcolumntype{L}[1]{>{\raggedright\let \newline\\\arraybackslash}p{#1}}
\begin{tabular}{L{0.175\textwidth}L{0.29\textwidth}L{0.18\textwidth}L{0.19\textwidth}L{0.16\textwidth}L{0.1\textwidth}}
\toprule \toprule
Process & Generator & ME Order & PDF & Parton Shower & Tune \\
\midrule
\multicolumn{6}{c}{Signal samples}\\
\midrule
$ZH, H\rightarrow \gamma\gamma_d$ & \POWHEGBOX[v2] & \NLO & \NNPDF[3.0nlo] & \PYTHIA[8.245] & AZNLO \\ 
\midrule
\multicolumn{6}{c}{SM background samples}\\
\midrule
$V\gamma^{QCD}$  & \sherpa~v2.2.8 & \NLO (up to 2 jets), \newline LO (up to 3 jets) & \NNPDF[3.0nnlo] & \Sherpa \MEPSatNLO & \Sherpa \\
$V\gamma^{EWK}$ & \mgamc[v2.6.5] & LO & \NNPDF[2.3lo] & \PYTHIA[8.240] &   A14 \\
$Z^{QCD}$ & \sherpa~v2.2.1 & NLO (up to 2 jets), \newline LO (up to 4 jets) &\NNPDF[3.0nnlo] & \sherpa \MEPSatNLO & \sherpa \\
$Z^{EWK}$ & \sherpa~v2.2.1 & NLO (up to 2 jets), \newline LO (up to 4 jets) &\NNPDF[3.0nnlo] & \sherpa \MEPSatNLO & \sherpa \\
Single $t$-quark/\ttbar & \POWHEGBOX[v2] & \NLO & \NNPDF[2.3lo] & \PYTHIA[8.230] &A14 \\
\ttbar ($V,VV$), $Wt\gamma$ & \mgamc[v2.2.3] & \NLO & \NNPDF[2.3lo] & \PYTHIA[8.210] & A14 \\
SM Higgs & \POWHEGBOX[v2] & \NNLO (ggF), NLO (VBF, $VH, t\bar{t}H$) & PDF4LHC15 & \PYTHIA[8.230] & AZNLO \\
$VV\gamma$ & \sherpa~v2.2.11 & \NLO (0 jets), \newline LO (up to 3 jets) & \NNPDF[3.0nnlo] & \Sherpa \MEPSatNLO & \Sherpa \\
$VV$/$VVV$ & \sherpa~v2.2.2 & \NLO (0 jets), \newline LO (up to 3 jets) & \NNPDF[3.0nnlo] & \Sherpa \MEPSatNLO & \Sherpa \\
\midrule
\multicolumn{6}{c}{Samples for evaluating systematic uncertainties}\\
\midrule
$ZH, H\rightarrow \gamma\gamma_{d}$ & \POWHEGBOX[v2] & \NLO & \NNPDF[3.0nlo] & \Herwig[v7.1.3] & H7-UE-MMHT \\
$Z\gamma^{QCD}$  & \mgamc[v2.3.3]  &\NLO & \NNPDF[2.3lo] & \PYTHIA[8.212] & A14 \\
$W\gamma^{QCD}$  & \mg[v2.8.1]  & \NLO & \NNPDF[2.3lo] & \PYTHIA[8.244] & A14 \\
$W\gamma^{EWK}$  & \mg[v2.8.1]  & \NLO & \NNPDF[3.0nlo] & \Herwig[v7.1.3] & H7-UE-MMHT  \\
$Z^{QCD}$  & \mgamc[v2.2.3]  &\NLO & \NNPDF[2.3lo] & \PYTHIA[8.210] & A14 \\
$Z^{EWK}$  & \Herwig[v7.1.3]  & \NLO & \NNPDF[3.0nlo] & \Herwig[v7.1.3] & H7-UE-MMHT  \\
$V\gamma\gamma$ & \mgamc[v2.7.3]  & \NLO & \NNPDF[2.3lo] & \PYTHIA[8.244] & A14 \\
\ttbar  &  \POWHEGBOX[v2] & \NLO & \NNPDF[3.0nlo] & \Herwig[v7.0.4] & H7-UE-MMHT  \\
\ttbar$V$  &  \mgamc[v2.3.3]  & \NLO & \NNPDF[3.0nlo] & \Herwig[v7.2.1] & H7-UE-MMHT  \\
\bottomrule \bottomrule
\end{tabular}
}
\end{table*}

\subsection{Signal samples}
\label{subsec:zhyyd}
The signal process of a Higgs boson decaying to a photon and an invisible dark photon ($\gamma_{d}$) was generated in the $ZH$ production mode as shown in Figure~\ref{fig:zhyyd}. Both $q\bar{q}\to ZH$ and $gg\to ZH$ production modes were considered in order to have the full $ZH$ cross-section of 0.884~pb~\cite{deFlorian:2016spz}. Matrix elements were estimatd using \POWHEG{}v2~\cite{Campbell:2012am,Luisoni:2013cuh} with the NNPDF3.0 parton density libraries~\cite{Ball:2014uwa}. Events from the $q\bar{q}\to ZH$ process were generated at next-to-leading order (NLO) while the ones corresponding to $gg\to ZH$ were generated at leading order (LO). \PYTHIA{}8.245 was used to perform the Higgs boson decay as well as the hadronisation and showering with the CTEQ6L1 PDF set and the AZNLO tune~\cite{Pumplin:2002vw}. The samples were generated at a Higgs boson mass of 125~\GeV{}, a width set to the SM value of 4~\MeV{}~\cite{deFlorian:2016spz} and the complex pole scheme~\cite{Goria:2011wa} turned off. The Hidden Valley scenario~\cite{Carloni_2010} for BSM Higgs boson decay as implemented in \PYTHIA{}8.150 was used to produce $H \to \gamma\gamma_d$ signal events. A total of six Monte Carlo samples were generated with dark photon masses equal to 0, 1, 10, 20, 30, 40~\GeV{}. Finally, in order to increase the generation efficiency, a generator-level lepton filter was applied, requiring two electrons or muons with $\pT > 10$~\GeV{} and $|\eta|<2.7$.


\subsection{Background samples}
\label{subsec:bkgMC}
 
The analysis is affected by a large variety of background processes.
The irreducible background comes from $VV\gamma$ final states ($V$ being any of $W,Z$) with both $V$ bosons decaying leptonically.
The reducible background, which is dominant, comes from  \met\ mismeasurement (fake \met) --- typically due to undetected particles or hadronic jets not fully contained in the detector acceptance --- or from particle misidentification. For instance, an electron can wrongly be identified as a photon (electron faking photon, $e\rightsquigarrow\gamma$), or an energetic neutral pion ($\pi^0\to\gamma\gamma$) contained in a hadronic jet can wrongly be identified as a photon (jet faking photon, $j\rightsquigarrow\gamma$). Moreover, a track in a jet may also be wrongly identified as a lepton, or a lepton from a heavy-flavour quark decay may appear as an isolated lepton (jet faking lepton, $j\rightsquigarrow\ell$). Finally, events from top-quark production, with subsequent semi-leptonic $t\to W(\to\ell\nu)b$ decay, contain genuine \met, one or two leptons, and one or two $b$-jets that may fail a $b$-jet veto.
 
Whenever higher-order cross-section computations are available~\cite{Melnikov:2006kv,LHCTopWGttbarXsec,LHCTopWGsgtopXsec}, they are used to rescale the cross-section of the generator. Finally, a specific treatment was applied to processes with possible overlaping events in $Z\gamma$ and $Z$+jets, $VV$ and $VV\gamma$ or in $t\bar{t}$ and $t\bar{t}\gamma$. In order to avoid duplicated events, the overlap removal algorithm gives preference to photons produced in matrix elements (ME) over the ones from intial/final state radiations or decays. As an example, overlapping events were removed from $Z$+jets and kept in the $Z\gamma$ process. Finally, $\ell\ell$ final states consist of $ee$ and $\mumu$ channels and include the leptonic decay of $\tau$-leptons.

\paragraph{Irreducible background:}
These processes were generated according to the number of leptons in their final states as $\ell\ell\nu\nu\gamma$, $\ell\ell\ell\nu\gamma$ and $\ell\ell\ell\ell\gamma$ using the \SHERPA~v2.2.11~\cite{Bothmann:2019yzt} event generator at NLO with up to three jets at the leading order (LO). In addition, contributions from off-shell bosons were also included. The full event was generated using NNPDF30\_nnlo\_as\_0118 libraries and tuning developed by the
Sherpa authors~\cite{Ball:2014uwa}.
 
\paragraph{\Zg and \Wg productions:}
The \Zllg{} and \Wlng{} processes are split into two components based on the
order in the electroweak coupling constant \alphaewk{}. The strongly-produced component is of order $\alpha^3_{\mathrm{EWK}}$ and the electroweak component is of order $\alpha^5_{\mathrm{EWK}}$. The strong background processes of \Zllg{} are modelled using filtered \SHERPA{}~2.2.8~\cite{Gleisberg:2008ta} samples. These samples are filtered by requiring a vector boson \pT{} $>$ 90~\GeV{} and are merged with \SHERPA{}~2.2.8 \Vg{} samples produced with a biased phase space enhancing $\pT{}(V)$ and $\pT(\gamma)$ at high values. A photon filter was applied, with \pT $>$ 7~\GeV{}, for all the merged \Zllg{} samples. These calculations use the Comix~\cite{Gleisberg:2008fv} and Open-Loops~\cite{Cascioli:2011va} matrix element generators, and merging is done with the \SHERPA{} parton shower~\cite{Schumann:2007mg} using the ME$+$PS@NLO prescription~\cite{Hoeche:2012yf}. The NNPDF30\_nnlo\_as\_0118 PDF and tuning~\cite{Ball:2014uwa} is used. Matrix elements for the strongly-produced contribution are calculated at NLO in \alphas{} for up to two additional final-state partons, and at LO for up to three additional partons with NLO EWK$+$QCD corrections used as the central value for \SHERPA{}. Matrix elements for the electroweak VBF contribution are calculated at LO in \alphas{} with five final-state partons (e.g. $\ell^+\ell^-\gamma jj$ QCD=0) in \MADGRAPH{}~2.6.5. The LO interference between this electroweak and strong production samples was generated, and its cross-section is about 5\% of the electroweak sample.
This value is taken into account as an uncertainty on the electroweak background contribution. The showering is done using \PYTHIA{}8.240~\cite{Sjostrand:2014zea} and merged in the CKKW-L scheme~\cite{L_nnblad_2012}. The photon \pT{} is required to be larger than 10~\GeV{}, and the Frixione isolation~\cite{Frixione:1998jh} is applied to remove overlap with charged partons.
 
\paragraph{\Zjets production:}
The modelling of $Z$+jets is crucial in this analysis. This background has significant systematic uncertainties, as the modelling of the \MET depends on the modelling of pile-up interactions and on the jet energy response. The $Z$+jets samples used are simulated using the \SHERPA{}~2.2.1 event generator and the NNPDF3.0 NNLO libraries and the \SHERPA{} tuning, with the invariant $\ell\ell$ mass $m_{\ell\ell}$ $> 40$ \GeV{}.
 
\paragraph{Top-quark pair and single top-quark productions:}
Background samples for top-quark pair production, as well as single top-quark, including $Wt$ production, were simulated using \POWHEGBOX[v2] interfaced with \PYTHIA{}8.230 with the NNPDF2.3LO libraries and the A14 tuning was used. Both $s$-channel and $t$-channel productions are included in the single top-quark samples. The diagram removal scheme (DR) \cite{Frixione:2008yi} is used to remove interferences and overlap between the $tW$ and $t\bar{t}$ productions.
 
\paragraph{$t\bar{t}$ ($V$, $VV$), $t\bar{t}\gamma$ and $Wt\gamma$ production:}
Background samples for top-quark pair production in association with one or two vector bosons ($W$ or $Z$) or a photon, as well as single top-quark with an additional $W$ and $\gamma$ were  simulated with the \MADGRAPH{}5\_aMC@NLO~2.3.3 generator~\cite{Alwall:2014hca} interfaced with \PYTHIA{}8.210 with the NNPDF2.3LO libraries and the A14 tune.
 
\paragraph{SM Higgs boson production:}
Different production modes are considered as backgrounds as they can produce ffinal states with $\ell^+\ell^-+\gamma+\MET{}$. Higgs bosons produced in association with a $W$ or $Z$ boson as well as Higgs bosons produced via gluon-gluon fusion (ggF), vector boson fusion (VBF) and $ttH$ productions were all considered. These samples were generated using \POWHEGBOX[v2] with the NNPDF3.0 libraries and then showered with \PYTHIA{}8.230 and AZNLO tuning.
 
\paragraph{Di-boson production:}
\zzllll , \zzllvv , \wzlvll and \wwlvlv ($\ell = e, \mu$) processes are simulated using the \SHERPA{}2.2.2~\cite{Bothmann:2019yzt} event generator with NNPFD3.0NNLO libraries in the case of $qq$ and $gg$-initiated production. The $gg\rightarrow ZZ$ processes include a QCD k-factor of 1.7 which was calculated from the ratio NLO/LO of the corresponding cross-sections at 13~\TeV{} \cite{deFlorian:2016spz}. The  \qqllvv samples include both $ZZ$ and $WW$ events.
 
\paragraph{Tri-boson production:}
The expected contribution from this background is very minor as it is suppressed by the requirement of no more than two leptons in the final state. Tri-boson production, $VVV$, with $V$ = $W$ or $Z$, is simulated by the \SHERPA{}2.2.2 event generator at NLO with NNPDF3.0 NNLO PDF libraires and tuning.



\section{Event reconstruction and selection}
\label{sec:reco_selec}

\subsection{Event reconstruction}
\label{subsec:reco}
 
Candidate events are required to have a reconstructed vertex with at least two associated tracks, originating from the beam collision region in the $x-y$ plane,  with $\pT{}>0.5$~\GeV{}. The primary vertex in the event is selected as the vertex with the highest scalar sum of the squared \pT of associated tracks \cite{ATL-PHYS-PUB-2015-026}. 
 
Electrons are reconstructed by matching clustered energy deposits in the EM calorimeters to tracks in the ID~\cite{EGAM-2018-01}. Candidates falling within the transition regions between the barrel and endcap EM calorimeters ($1.37 < |\eta| < 1.52$) are also included. All electrons must fulfill \pT$> 4.5$~\GeV{} and $|\eta| < 2.47$ as well as loose identification criteria~\cite{EGAM-2018-01}. A longitudinal impact parameter requirement of $|z_0\sin(\theta)|<5$~mm as well as a transverse impact parameter satisfying $|d_0|/\sigma(d_0)<5$ are required to ensure that electrons originate from the primary vertex.
 
Muons are reconstructed by matching ID tracks to MS tracks or track segments, by matching ID tracks to a calorimeter energy deposit compatible with a minimum-ionizing particle, or by identifying MS tracks passing the loose requirement and compatible with the primary vertex~\cite{MUON-2018-03}. Muon candidates are required to fulfill $\pT >4$~\GeV{}, $|\eta|<2.7$ and a very loose isolation criterion. A longitudinal impact parameter requirement of $|z_0\sin(\theta)|<5$~mm as well as a transverse impact parameter satisfying $|d_0|/\sigma(d_0)<3$ are required to ensure that muons orginate from the primary vertex as well.
 
Photon candidates are reconstructed from clustered energy deposits in the EM calorimeter \cite{EGAM-2018-01}. They must fulfill \pT$>15$~\GeV{}, $|\eta| < 2.37$ and tight identification and isolation criteria~\cite{EGAM-2018-01}. They are also required to be outside the transition region ($1.37< |\eta| < 1.52$) between barrel and endcap EM calorimeters.
 
Particle flow (PFlow) jets are reconstructed using the anti-$k_t$ algorithm \cite{Cacciari:2008gp,Fastjet} with a radius parameter of $R = 0.4$, using charged constituents associated with the primary vertex and neutral PFlow constituents as inputs \cite{PERF-2015-09}. A complete energy calibration procedure, which recovers the initial parton energies after removing pile-up effects, is then applied \cite{JETM-2018-05}. Jet candidates are required to have \pT$>$ 20~\GeV{} and $|\eta|< 4.5$. A jet vertex tagger (JVT) discriminant \cite{ATLAS-CONF-2014-018} is used to identify jets within $|\eta|<2.5$ originating from the hard scattered (HS) interaction through the use of tracking and vertexing. For jets with $|\eta|>2.5$, a forward jet vertex tagging algorithm (fJVT) \cite{ATL-PHYS-PUB-2015-034,ATL-PHYS-PUB-2019-026} is used to reject pile-up jets. The working points used for JVT (fJVT) provide an HS selection efficiency of about 97\% (93\% for \pT$>50$~\GeV). In the case of $b$-jets, a multivariate discriminant output is used for the identification \cite{PERF-2016-05}. The used working point provides a 77\% $b$-jet tagging efficiency in simulated inclusive $t\bar{t}$ events, with rejection factors of 6 and 134 for charm-hadron jets and light-flavour quark- or gluon-initiated jets, respectively.
 
The missing transverse momentum \METvec{}, originating from the unbalanced momentum in the transverse plane, is defined as the negative vectorial sum of the transverse momenta of all selected (hard objects) electrons, muons, photons, PFlow jets, as well as the ``soft term'', which is estimated from tracks compatible with the primary vertex, but not matched to any of those objects \cite{PERF-2016-07}. Jets are only included in the \METvec{} definition if they have $\pT{}>20$~GeV and satisfy the tight JVT and fJVT selections in order to mitigate pile-up effects. Moreover, an \MET{} significance $S$ is defined to reduce the effects on \MET{} from resolution fluctuations.$S$ is a powerful quantity to discriminate between events with fake \METvec, arising from instrumental sources or poorly reconstructed physics objects and events with genuine  \METvec originating from weakly interacting particles, like neutrinos. The variable $S$ is calculated as |\,\METvec\,|$/\left[\sigma_\text{L}^2\left(1-\rho_\text{LT}^2 \right) \right]^{1/2}$, where $\sigma_\text{L}$ is the total standard deviation in the direction longitudinal to the \METvec\ corresponding to the summation of the covariance matrices from resolution effects of physics objects and  soft term entering the \METvec\ calculation and $\rho_\text{LT}$ is the correlation factor of the longitudinal (L) and transverse (T) measurements~\cite{ATLAS-CONF-2018-038}.
 
To resolve ambiguities in the reconstruction of physics objetcs and to avoid double counting of energy deposits and momentum measurements, a \DeltaR{} separation is required as detailed in Table~\ref{tab:overlap_removal}.
 
\begin{table}[!htbp]
\caption{Overview of the overlap removal between reconstructed physics objects and the corresponding matching criteria in order of priority.}
\label{tab:overlap_removal}
\centering
\begin{tabular}{ccc}
\toprule \toprule
Remove & Keep & Matching criteria \\
\hline
jet & electron & $\DeltaR < 0.2$ \\
jet & muon & number of tracks $<$ 3 and $\DeltaR < 0.2$ \\
jet & photon & $\DeltaR < 0.4$ \\
electron & jet & $ 0.2 < \DeltaR < 0.4 $ \\
electron & muon & shared same ID track \\
electron & electron & shared same ID track, electron with lower \pT removed \\
muon & jet & 0.2 < $\DeltaR < 0.4$ \\
muon & electron & muon with calorimeter deposits and shared ID track \\
photon & electron & $\DeltaR < 0.4$ \\
photon & muon & $\DeltaR < 0.4$ \\
\bottomrule \bottomrule
\end{tabular}
\end{table}
 
Finally, several cleaning requirements are applied to suppress non-collision backgrounds~\cite{ATLAS-CONF-2015-029}. Misreconstructed (bad quality) jets can be caused by electronic noise, and jets from collisions are identified by requiring a good fit to the expected pulse shape for each constituent calorimeter cell. Cosmic-ray showers and beam-halo interactions with the LHC collimators are another source of misreconstructed jets. Those jets are identified by requirements on their energy distribution in the calorimeter and the fraction of their constituent tracks that originate from the primary vertex. Events are rejected if they contain a bad quality jet with $\pT > 20$~\GeV{} as they tend to give rise to fake \MET{} and a poor description of its distribution in the tail region.


 
\subsection{Signal region selection}
\label{subsec:evtsel}
 
The signal region (SR) targets the $\dil$+$\gamma$+\MET final state, where the two same-flavour, oppositely charged leptons come from a $Z$ boson decay, the photon comes from the Higgs boson decay and the \MET arises from the potential undetected dark photon. Accepted leptons are required to match the corresponding online trigger candidates, and trigger scale factors and their uncertainties are applied.
 
Table~\ref{tab:EventSelection} lists the criteria to select the events in the SR, optimised using a multivariate approach to define the rectangular selection to maximize the signal over background acceptance. A BR($H\to \gamma\gamma_d $)~=~5\% was assumed for all dark photon masses. The resulting acceptance times efficiency for combined $qqZH$ and $ggZH$ signal events amounts to 10.5\% for all considered $\gamma_d$ masses.
\begin{table}[h]
\begin{center}
\caption{Optimised kinematic selections defining the signal region for $\dil$+$\gamma$+\MET.}
\label{tab:EventSelection}
\renewcommand\arraystretch{1.8}
\begin{tabular}{c}
\toprule
\toprule
\vtop{\hbox{\strut Two same flavour, opposite sign, medium ID and loose isolated leptons,}\hbox{\strut \hspace{1.5cm} with leading $\pT{}>27$~\GeV{}, sub-leading $\pT{}>20$~\GeV{} }} \\
\hline
Veto events with additional lepton(s) with loose ID and $\pT\,>$ 10 \GeV{} \\
\hline
$76~\GeV{} < m_{\ell\ell} < 116$~\GeV{}\\
\hline
Only one tight ID, tight isolation photon with $\ET^{\gamma}>25$~\GeV{} \\
\hline
$\met{} > 60$~\GeV{} with $\DphiEtmisslly>2.4~\mathrm{rad}$ \\
\hline
$m_{\ell\ell\gamma} > 100$~\GeV{} \\
\hline
$\text{N}_{\text{jet}} \leq 2$, with $\pT^{\text{jet}} > 30~\GeV{}$, $|\eta|<4.5$\\
\hline
Veto events with $b$-jet(s) \\
\bottomrule
\bottomrule
\end{tabular}
\end{center}
\end{table}
 
Accepted signal electrons, reconstructed using a likelihood-based identification algorithm, must fulfill a `Medium' criteria~\cite{EGAM-2018-01}, while `Loose' criteria are used to veto additional electrons in the SR. The likelihood relies on the shape of the EM shower measured in the calorimeter, the quality of the track reconstruction, and the quality of the match between the track and the cluster. Similarly, selected muons are also required to have `Medium' quality identification \cite{MUON-2018-03} while `Loose' criteria are used to veto additional muons. To suppress hadronic and non-prompt lepton backgrounds, electron and muon candidates are required to satisfy the particle-flow `Loose' isolation criteria, which are based on tracking and calorimeter measurements~\cite{MUON-2018-03}. The veto on events with a third loose lepton is used to reduce \zzllll{} and \wzlvll{} contributions. The invariant mass of the selected lepton pairs $m_{\ell\ell}$ is required to be within the $Z$ boson peak ($[76-116]~\GeV{}$) to reject processes such as $t\overline{t}$ and \wwlvlv.
 
Exactly one `Tight' photon, satisfying the tight isolation criteria~\cite{EGAM-2018-01} is required. The isolation section requires that the sum of the transverse energies (at the electromagnetic scale) of positive-energy topological clusters located within a distance $\DeltaR{}=0.4$ of the photon candidate $\gamma$ must be less than $0.22\times E^{\gamma}_{{\mathrm T}} + 2.45$~[\GeV{}], where $E^{\gamma}_{{\mathrm T}}$ is the transverse energy of the photon. In addition, the scalar sum of the \pT of all tracks located within a distance $\DeltaR{}=0.2$ of the photon candidate must be less than $0.05\times E^{\gamma}_{{\mathrm T}}$~[\GeV{}].
 
Other selections are applied, which exploit the topology and kinematics of the signal events. A threshold on \MET was optimised to select signal events while rejecting the inclusive \Zboson production. In addition, \METvec{} is expected to be back-to-back with the $(Z-\gamma)$ system, leading to a requirement on the azimuthal separation \DphiEtmisslly\  applied in the SR. Furthermore, events with more than 2~jets (with $\pT^{jet}>30$~\GeV{}, $|\eta|<4.5$) are rejected to reduce contributions from $V$+jets processes. Finally, a veto on any $b$-tagged jet is also applied to reduce processes with a top quark.
 
Events are categorised into two sub-regions called, respectively, \ee- and \mumu-channel. The expected signal and background composition in each SR, as predicted from MC simulation, after all the optimisations, is shown in Table~\ref{tab:sig_bkg_SR}. It should be noticed that the $VV$ background is dominated by the $WZ$ process where the photon is the result of an electron mis-identification while contributions from $WW,ZZ$ are found to be negligible due to the very low probability of jets to be mis-identified as photons. The $VV\gamma$ background is dominated by $ZZ\gamma$ and $WW\gamma$ contributions from the $\ell\ell\nu\nu\gamma$ final state.
 
\begin{table}[h]
\caption{Expected event yields for signal and background in the SR corresponding to $\mathcal{L}=139$~fb$^{-1}$. Signal events are for massless $\gamma_d$, assuming BR($H\to \gamma\gamma_d) = 5\%$. Events for background processes  are categorised as $Z\gamma$ (QCD+EWK $Z\gamma$), $Z+$jets (QCD+EWK $Z+$jets), Top (single top-quark, $Wt$), $t\bar{t}$ ($t\bar{t}, t\bar{t}V, t\bar{t}VV$), Top$\gamma$ ($Wt\gamma$), $VV\gamma$ ($WW\gamma, WZ\gamma, ZZ\gamma$), $VV$ ($WW, WZ, ZZ$), SM Higgs ($ggH, VH$, VBF $H$) and $W\gamma$ (QCD+EWK $W\gamma$). Only statistical uncertainties on the simulated samples are shown.}
\centering
\resizebox{\columnwidth}{!}{
\begin{tabular}{cccccccccccc}
\toprule
\toprule
Channel     &     Signal  &    $Z\gamma$     &     $Z+$jets     &     $Top$     &     $t\bar{t}$     &     Top$\gamma$     &     $VV\gamma$     &     $VV$     &     SM Higgs      &     $W\gamma$     &     Total background     \\
\midrule
$ee$          &    $19.3 \pm 0.2$      &     $155 \pm 15$      &     $274 \pm  55$      &   $3.5 \pm 0.7$   &   $25 \pm  1$  &  $1.9 \pm 0.1$      & $26 \pm 1 $    &  $27 \pm 1$   & $0.41 \pm 0.01$    & $3.5 \pm 1.5$   & $517 \pm 57$        \\
\midrule
 
$\mu\mu$           &    $22.4 \pm 0.2$      &     $283 \pm 18$      &     $380 \pm  63$      &   $4.6 \pm 0.8$   &   $26 \pm  1$  &  $2.4 \pm 0.1$   & $35 \pm 1 $    &  $24 \pm 1$   & $0.54 \pm 0.01$    & $1.6 \pm 1.1$   & $758 \pm 66$        \\
 
\bottomrule
\bottomrule
\end{tabular}
}
\label{tab:sig_bkg_SR}
\end{table}
 
In order to enhance the sensitivity of the search for the $\gamma_d$ signal, a boosted decision tree (BDT) algorithm was implemented using the XGBoost classifier~\cite{XGBoost}. For training and testing, all events entering the SR are used. All signal events are assigned to the positive class (y = 1) and all background events are assigned to the negative class (y = 0) in the training data. The model has been trained using a feature set consisting of the following 6 variables, according to their ranking: $S$ (\MET significance), $m_{\text T}$ (Equation~\ref{eq:MT}),  $m_{\ell\ell}$,  $\pT^{\gamma}$, $m_{\ell\ell\gamma}$,  and $\pT^{ratio}$ (Equation~\ref{eq:abspt}), with:
 
\begin{equation}
m_{\text T} = \sqrt{2\MET{}\pT^{\gamma}[1-\cos[\Delta \phi (\METvec{},\Vec{p}_{\text T}^{\gamma})]]}
\label{eq:MT}
\end{equation}
 
\begin{equation}
\pT^{ratio}  =\frac{|\METvec{}+\Vec{p}_{\text T}^{\gamma}|-\pT^{\ell\ell}}{\pT^{\ell\ell}}
\label{eq:abspt}
\end{equation}
 
To train the BDT, the simulated samples introduced in section~\ref{sec:sim} for the signal and the background processes are used. The signal samples for all considered $\gamma_d$ masses were merged and the same weight was assigned to their events. To enhance the statistical power of the training sample, a five-fold cross validation strategy was adopted. The simulation samples were divided into five equal subsets. Five BDT classifiers with the same input variables were trained. In each BDT training, four subsets of simulation samples were used as the training sample, and the remaining subset was used as the testing sample.  The five BDT trainings correspond to the five possible permutations of such training-testing setups. The five trained BDT models were used to calculate the BDT score of data that were divided into five subsets in the same manner. The BDT classifier output is then used as an observable for the final statistical analysis as discussed in Section~\ref{sec:results}.


\section{Treatment of the background processes}
\label{sec:bkg}

 
The different background sources, described in Section~\ref{subsec:bkgMC}, are classified into disjoint categories, described below, according to their treatment in the analysis. In what follows, weak bosons $W,~Z$ are always meant to decay leptonically: $W\to\ell\nu$ and $Z\to\dil$, unless stated otherwise.
 
\begin{itemize}
\item \textbf{Irreducible background from $VV\gamma$ final states}: $Z(\to\dil)Z(\to\nunu)\gamma$ and $W^+(\to\ell^+\nu)W^-(\to\ell^-\bar{\nu})\gamma$~.
\item \textbf{Background from electrons faking photons} ($e\rightsquigarrow\gamma$): the involved final states are mostly \mbox{$Z(\to\dil)W(\to e\nu)$}, and to a lesser extent also $Z(\to\dil)Z(\to\ee)$ with an undetected electron, $VVV$ with undetected leptons, and finally $\dil Vt$ and $VV\ttbar$ that were not
rejected by the $b$-veto.
\item \textbf{Background from fake \met}: the largest contribution comes from $Z\gamma+\mathrm{jets}$ and $Z+\mathrm{jets}$, where the \met mismeasurement is mostly due to jet energy mismeasurement. In order for the $Z+\mathrm{jets}$ to mimic the signal sought, one jet must be misreconstructed as a photon;
this source of background is therefore included in the fake \met estimation.
Minor contributions to this category come from $Z\gamma\gamma$ and $ZH(\to\gamma\gamma)$ with an undetected photon, and from $gg\to H$
and VBF 
Higgs production with subsequent $H\to Z\gamma$ decay. The $ZZ$ and $VVV$ final states are not included in this category, as they are already accounted for in the $e\rightsquigarrow\gamma$ case.
\item \textbf{Background from top quark production}: the main final states are $tW\gamma$, $\ttbar\gamma$, $\ttbar$ and single top-quark production --- in the last two cases, the identified photon comes from $j\rightsquigarrow\gamma$. The $\dil Vt$ and $VV\ttbar$ events are not included in this category, as they already enter the $e\rightsquigarrow\gamma$ category.
\item \textbf{Background from $W\gamma$}: enters the SR due to a jet faking a lepton.
\item \textbf{Other background from Higgs: $\ttbar H(\to Z\gamma)$ and $VH(\to Z\gamma)$}: this category can be divided into $\ttbar H(\to Z\gamma)$, $ZH(\to Z\gamma)$ and $W(\to\ell\nu)H(\to Z\gamma)$. Their experimental effects differ: the first one fails the $b$-veto; the second is irreducible, if one of the two $Z$s decays to $\nunu$ and the other to $\dil$; the third has genuine \met\ and an undetected lepton. However, their contributions are so tiny (by 3 or 4 orders of magnitude) with respect to any other concurrent process that it is more convenient to treat them all in a dedicated category. Note that this category does not include all processes involving the Higgs boson, as some are already accounted for in previously described categories.
\end{itemize}
 
The background contributions coming from the fake \met\ category (which is dominant) and from the $e\rightsquigarrow\gamma$ category are estimated with data-driven techniques. The other categories are estimated from MC simulation. The normalization for the top-quark background is checked in a dedicated validation region (VR), while the normalization of $VV\gamma$ irreducible background is adjusted in a dedicated control region (CR), as described in Section~\ref{sec:bkg_top_irred}.
 
\subsection{Evaluation of the background from electrons faking photons}
\label{sec:bkg_elefake}
 
This background occurs because electrons and photons produce similar shower shapes in the EM calorimeter, and photons may convert early in the tracker into an asymmetric \ee\ pair where one track may not be reconstructed; to keep the photon detection efficient, such cases must be considered, at the price of having a background from electrons wrongly identified as photons.
 
As this background cannot be perfectly modeled in simulation, a two step data-driven approach has been performed. First, the ``electron-to-photon fake rate'', $\Fey$ is estimated. This rate is then applied to a ``probe-electron'' CR, whose events are selected with the same criteria as for the SR, but requiring an extra electron instead of the photon --- called ``probe electron'', ${\eprobe}$, in the following.
 
To estimate \Fey, two set of data events are selected, respectively with $\ee$ and $e^\pm\gamma$ final states, with no additional leptons or photons. The distributions of the invariant masses, \mee\ and \mey, for the two final states, both exhibit a peak around the $Z$ mass, on top of a continuous background. The $Z$ contribution is extracted by means of a functional fit, where the peak is modeled as a double-sided Crystal Ball function, while the continuum is described by an exponential of a polynomial function. The quantity \Fey is computed from the results of this fit as the ratio of the number of events from $Z\to\ee$ and from $Z\rightsquigarrow e^\pm\gamma$ where the second category counts the $e^\pm\gamma$ mis-identification.
 
The estimate is done in several bins of photon \pT\ and $\eta$, as \Fey\ exhibits a strong dependance on these two variables. The values of \Fey\ increase both as a function of $\eta$ and \pT, ranging from 1.3--3\% in the barrel, to about 7\% at high \pT\ in the endcap. The overall uncertainty on each value is evaluated combining in quadrature the statistical uncertainties of the data samples, a systematic uncertainty quantifying the deviation between the actual and fitted peak, and an additional systematic uncertainty from a non-closure of the whole procedure carried out on MC simulation. The overall relative uncertainty on \Fey\ ranges from 7\% to 11\%.
 
The probe-electron CR (${\eprobe}$-CR) is populated with $\dil{\eprobe}+\met$ events, where the probe electron must satisfy the same kinematic requirements as the photon: in particular $\pT^{{\eprobe}}>25~\gev$, $|\eta^{{\eprobe}}|<1.37$ or $|\eta^{{\eprobe}}|\in[1.52,2.37]$, $m_{\ell\ell{\eprobe}}>100~\gev$, and $\dphiEtmisslle>2.4~\mathrm{rad}$. The rest of the event satisfies the same requirements as for the SR.
Each event in the ${\eprobe}$-CR is then scaled by the respective \Fey, computed as a function of $\eta^{{\eprobe}}$ and $\pT^{{\eprobe}}$; the number of the scaled events gives the estimate of the $e\rightsquigarrow\gamma$ background in the SR. Since this procedure is done per-event, it also allows for the estimation of the distributions of all kinematic variables for this background.
 
For the \mumu channel and the \ee channel the final states of the ${\eprobe}$-CRs are respectively $\mumu{\eprobe}+\met$ and $\ee{\eprobe}+\met$; in the latter case, in principle, any of the three electrons could be mis-identified as a photon and considered as the ${\eprobe}$ but the opposite-charge requirement on the $\ee$ pairs rules out one of the three possibilities.

In the ${\eprobe}$-CR, there is a contamination of jets faking electrons, occurring at a rate $f_{j\rightsquigarrow e}$, which is estimated from simulation at the level of 6\%--7\%. This is accounted for by rescaling all events in the ${\eprobe}$-CR by a quantity $1-f_{j\rightsquigarrow e}$; a systematic uncertainty equal to $f_{j\rightsquigarrow e}$ is assigned to this correction.
 
The data-driven estimates of the background from $e\rightsquigarrow\gamma$ in the \ee channel and \mumu channel are, respectively, $21.0\pm2.4$ and $20.4\pm2.1$ events. The total uncertainty is evaluated combining in quadrature the statistical uncertainty of the ${\eprobe}$-CR, the uncertainty on \Fey\ and on $f_{j\rightsquigarrow e}$.

\subsection{Evaluation of the background from fake \met}
\label{sec:bkg_fakemet}
 
\begin{figure}[h]
\centering
\includegraphics[width=0.5\columnwidth]{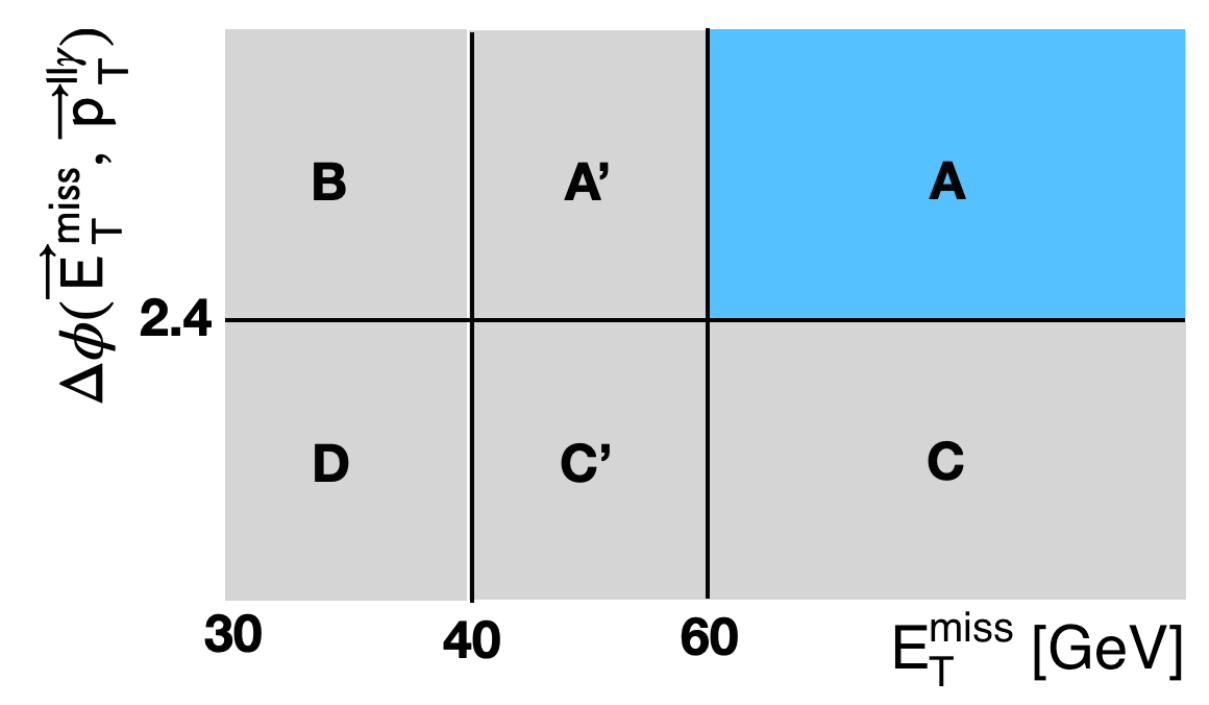}
\caption{Regions involved in the ``ABCD'' method for estimation of the background from fake \met; region $A$ is the signal region.}
\label{fig:bkg_fakemet_ABCD}
\end{figure}

The data-driven estimate of the background from fake \met\ is achieved by means of an ``ABCD method''; besides the SR (here labelled region $A$), three CRs ($B,~C,~D$) are defined by inverting the selection cuts on the \met\ and \dphiEtmisslly\ variables:
 
\begin{itemize}
\item \textbf{region $A$}: $\met>60~\gev$ and $\dphiEtmisslly>2.4~\mathrm{rad}$;
\item \textbf{region $B$}: $\met\in[30,40]~\gev$ and $\dphiEtmisslly>2.4~\mathrm{rad}$;
\item \textbf{region $C$}: $\met>60~\gev$ and $\dphiEtmisslly<2.4~\mathrm{rad}$;
\item \textbf{region $D$}: $\met\in[30,40]~\gev$ and $\dphiEtmisslly<2.4~\mathrm{rad}$.
\end{itemize}
Two more validation regions (VR) are introduced:
\begin{itemize}
\item \textbf{region $A'$}: $\met\in[40,60]~\gev$ and $\dphiEtmisslly>2.4~\mathrm{rad}$;
\item \textbf{region $C'$}: $\met\in[40,60]~\gev$ and $\dphiEtmisslly<2.4~\mathrm{rad}$.
\end{itemize}
as illustrated in Figure~\ref{fig:bkg_fakemet_ABCD}.
 
The regions $B,~C,~D$ are built such as to be  enriched in events with fake \met. The residual contribution of events with genuine \met\ is subtracted; a large portion of them come from the $e\rightsquigarrow\gamma$ background, and therefore can be evaluated and subtracted  by applying the data-driven procedure described in Section~\ref{sec:bkg_elefake} to the specific regions. The other processes ($VV\gamma$, top, $W\gamma$, and $\ttbar H+VH$) are estimated from simulation and subtracted. Then, the fake \met\ background in region $A$ can be computed as:
\begin{equation}
N_{~\!\!\!A}^{\fakeMET} = R \cdot \frac{~N_{~\!\!\!B}^\fakeMET \cdot N_{~\!\!\!C}^\fakeMET~}{N_{~\!\!\!D}^\fakeMET}
\label{eq:bkg_fakemet_R}
\end{equation}
where $R$ is a parameter that accounts for the correlation between the variables \met\ and \dphiEtmisslly\ in the fake \met\ processes --- if the two variables were independent, $R=1$. The variables were chosen so as to have $R$  close to 1 and to be stable when varying the requirements on each of the variables used.
 
The value of $R$ is estimated from MC simulation; the dominant processes to the fake \met\ background are $Z\gamma+\mathrm{jets}$ and $Z+\mathrm{jets}$.
In these processes, only few events enter the $A,~C$ regions, causing the statistical uncertainty on $R$ to be large; to overcome this, regions $(A+A')$ and $(C+C')$ are used instead, exploiting the stability of $R$ with respect to the value of the \met\ cut. Moreover, the compatibility of the $R$-value obtained with this approach with that from $A,~C$ regions has been checked. The extracted value for:
 
\begin{equation}
R_\mathrm{MC}=\frac{N_{~\!\!\!A+A'}^{Z(\gamma)+\mathrm{jets}}\cdot N_{~\!\!\!D}^{Z(\gamma)+\mathrm{jets}}}{N_{~\!\!\!B}^{Z(\gamma)+\mathrm{jets}}\cdot N_{~\!\!\!C+C'}^{Z(\gamma)+\mathrm{jets}}}
\label{eq:bkg_fakemet_R_MC}
\end{equation}
 
is displayed in Table~\ref{tab:bkg_fakemet_R}, for the $ee$ and $\mu\mu$ channels; the quoted uncertainties are due to the statistical uncertainties on the MC simulation. The relative contribution of $Z\gamma+\mathrm{jets}$ and $Z+\mathrm{jets}$ processes, and the $j\rightsquigarrow\gamma$ fake rate occurring in the second case, are known not to be precisely modeled by MC simulation; for this reason, the contribution of the $Z+\mathrm{jets}$ process has been changed by $\pm50\%$ of the MC prediction, to check its impact on $R_\mathrm{MC}$. The result is also shown in Table~\ref{tab:bkg_fakemet_R}; the small variations are well covered by the statistical uncertainties, therefore there is no evidence that the amount of $Z+\mathrm{jets}$ has an impact on $R_\mathrm{MC}$.
 
A further check on the reliability of $R_\mathrm{MC}$ has been carried out exploiting the two VRs $A',~C'$, where a similar factor:
 
\begin{equation}
R'=\frac{N_{~\!\!\!A'}^{\fakeMET}\cdot N_{~\!\!\!D}^{\fakeMET}}{N_{~\!\!\!B}^{\fakeMET}\cdot N_{~\!\!\!C'}^{\fakeMET}}
\label{eq:bkg_fakemet_Rprime}
\end{equation}
 
is introduced, and computed from MC simulation and data. $R'_\mathrm{MC}$ is computed using $Z\gamma+\mathrm{jets}$ and $Z+\mathrm{jets}$ processes, while $R'_\mathrm{data}$ uses event counts after the subtraction of $e\rightsquigarrow\gamma$, $VV\gamma$, top, $W\gamma$ and $VH$ background processes. The results are shown in Table~\ref{tab:bkg_fakemet_R}; the comparison of $R'_\mathrm{MC}$ and $R'_\mathrm{data}$ shows no significant discrepancy. As a conclusion, the evaluation of $R_\mathrm{MC}$ is considered reliable, and is used in Equation~(\ref{eq:bkg_fakemet_R}).
 
\begin{table}[h]
\centering
\caption{
Values of $R$ for the $ee$ and $\mu\mu$ channels. $R_{\textnormal{MC}}$ is computed from simulation, assuming a relative amount of $Z+\mathrm{jets}$ with respect to $Z\gamma+\mathrm{jets}$ processes as predicted by the simulation (nominal), or changed by factors $0.5$ or $1.5$. $R'$ is computed using regions $A',C'$ instead of $(A+A')$ and $(C+C')$, to allow a comparison between simulation and data in a sample enriched by fake \met. Statistical uncertainties on the simulated samples are reported.
}
\begin{tabular}{c|c|cc|cc}
\toprule
\toprule
channel & \multicolumn{3}{c|}{$R_\mathrm{MC}$} & \multicolumn{2}{c}{$R'$}\\
\cline{2-6}
& nominal & $0.5 \times (Z+\mathrm{jets})$ & $1.5 \times (Z+\mathrm{jets})$ & $R'_\mathrm{MC}$ & $R'_\mathrm{data}$ \\
\hline
$ee$     & $1.12\pm0.11$ & $1.15\pm0.13$ & $1.06\pm0.08$ & $1.09\pm0.11$ & $1.16\pm0.06$ \\
$\mu\mu$ & $1.24\pm0.11$ & $1.25\pm0.14$ & $1.23\pm0.09$ & $1.15\pm0.11$ & $1.18\pm0.05$ \\
\bottomrule
\bottomrule
\end{tabular}
\label{tab:bkg_fakemet_R}
\end{table}
 
The data-driven estimates of the fake \met\ background in the $ee$ and $\mu\mu$ channels are, respectively, $413\pm50$ and $581\pm64$ events. The errors are evaluated from the propagation of the statistical uncertainties of  $R_{\textnormal{MC}}$ and of data in the ABCD regions.
 
\subsection{Treatment of the irreducible background and the top-quark background}
\label{sec:bkg_top_irred}
 
The irreducible $VV\gamma$ background plays an important role at high values of the BDT score, which drive the sensitivity of the search; for this reason, a dedicated ``$VV\gamma$-CR'' is introduced, to correct the normalisation of this background (see Section~\ref{sec:results}). Recalling that such a background is a pure electroweak process, to which $Z(\to\dil)Z(\to\nunu)\gamma$ and $W^+(\to\ell^+\nu)W^-(\to\ell^-\bar{\nu})\gamma$ contribute, the ``$VV\gamma$-CR'' has been built to be enriched in $Z(\to\mumu)W(\to\mu\nu)\gamma$, thus requiring exactly three muons, one photon and no electrons. The opposite-charge \mumu-pair whose invariant mass is closest to the $Z$ mass must fulfil all $\dil$ kinematic cuts of the SR;  to reduce the statistical uncertainty, no requirements are applied on \met\ and \dphiEtmisslly. The purity of $ZW\gamma$ events in such a CR is estimated to be $83\%$ from simulation.
 
To check the normalization of the top-induced background predicted by simulation, a ``top-VR'' has been used, enriched by such processes; it is defined starting from the selection of the SR, removing the requirements on $m_{\ell\ell}$ and $\dphiEtmisslly$, and requiring at least one $b$-tagged jet. The purity of events with top-quark production is estimated to be $92\%$ from simulation in such a VR.
 
\subsection{Background checks in validation region}
 
\begin{figure}[h]
\centering
\subfloat[]{\includegraphics[scale=0.37]{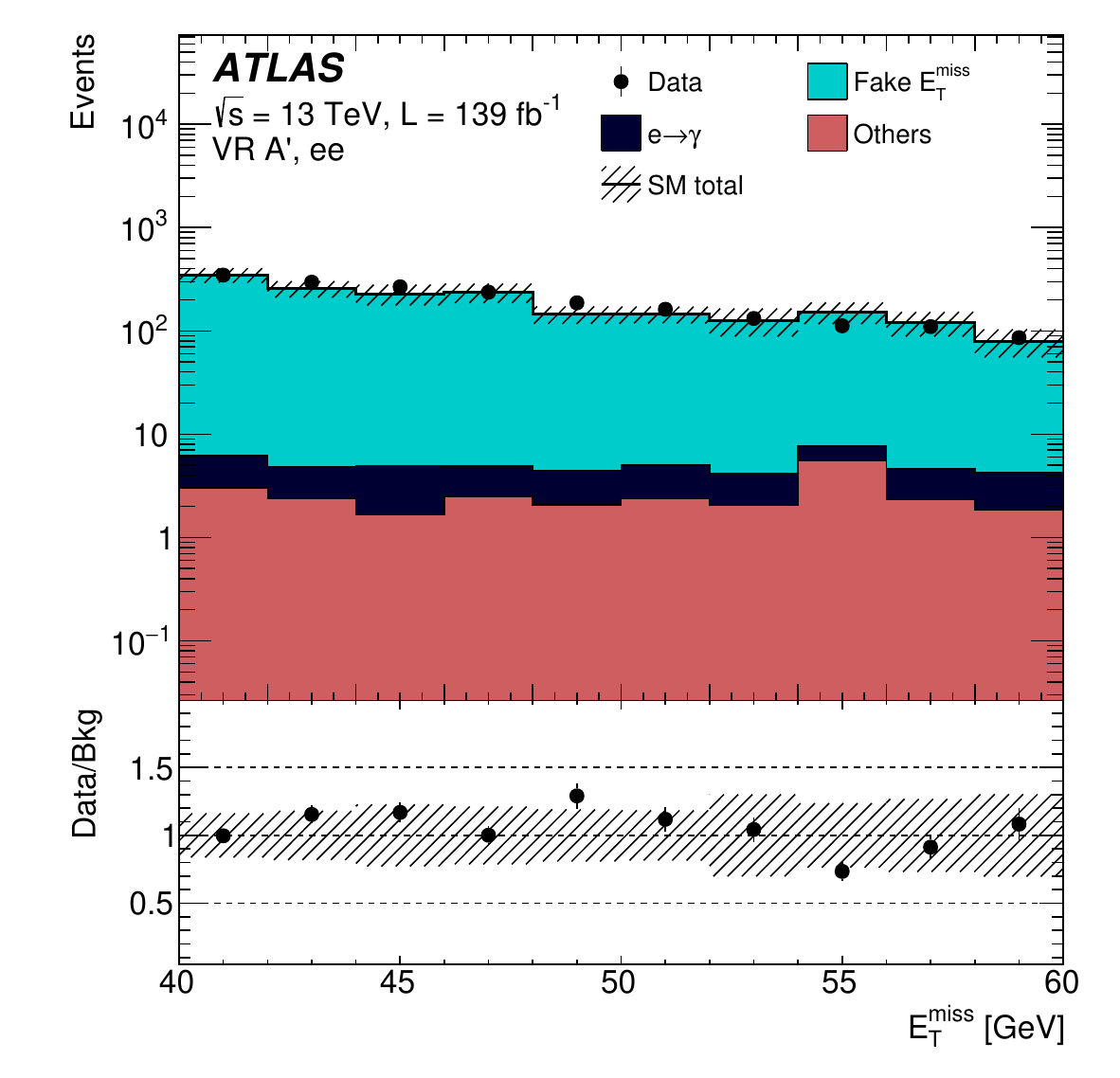}}
\subfloat[]{\includegraphics[scale=0.37]{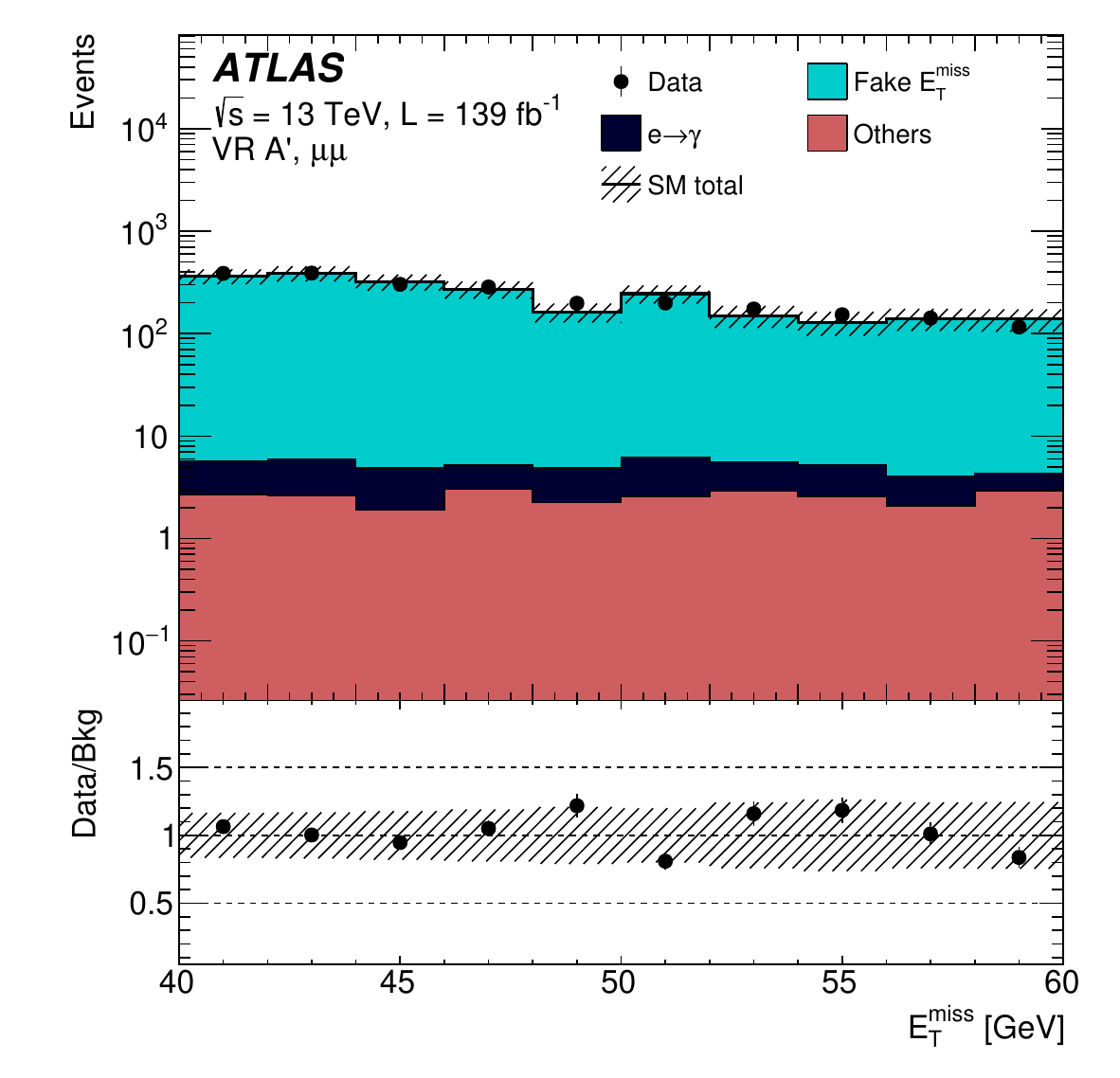}}
\caption{
Comparison between the expected background and data in the validation region $A'$, as a function of \met, for the $ee$ (a) and $\mu\mu$ (b) channels. The background yields from ``fake \met'' and ``$e\rightsquigarrow\gamma$'' are estimated with data-driven techniques. The other backgrounds are obtained from simulation and have been merged together. Uncertainties shown are statistical, both for data and for simulated backgrounds, while for the data-driven backgrounds the systematic uncertainties related to the method are also included.
}
\label{fig:bkg-Aprime-ETmiss}
\end{figure}

\begin{figure}[h]
\centering
\subfloat[]{\includegraphics[scale=0.37]{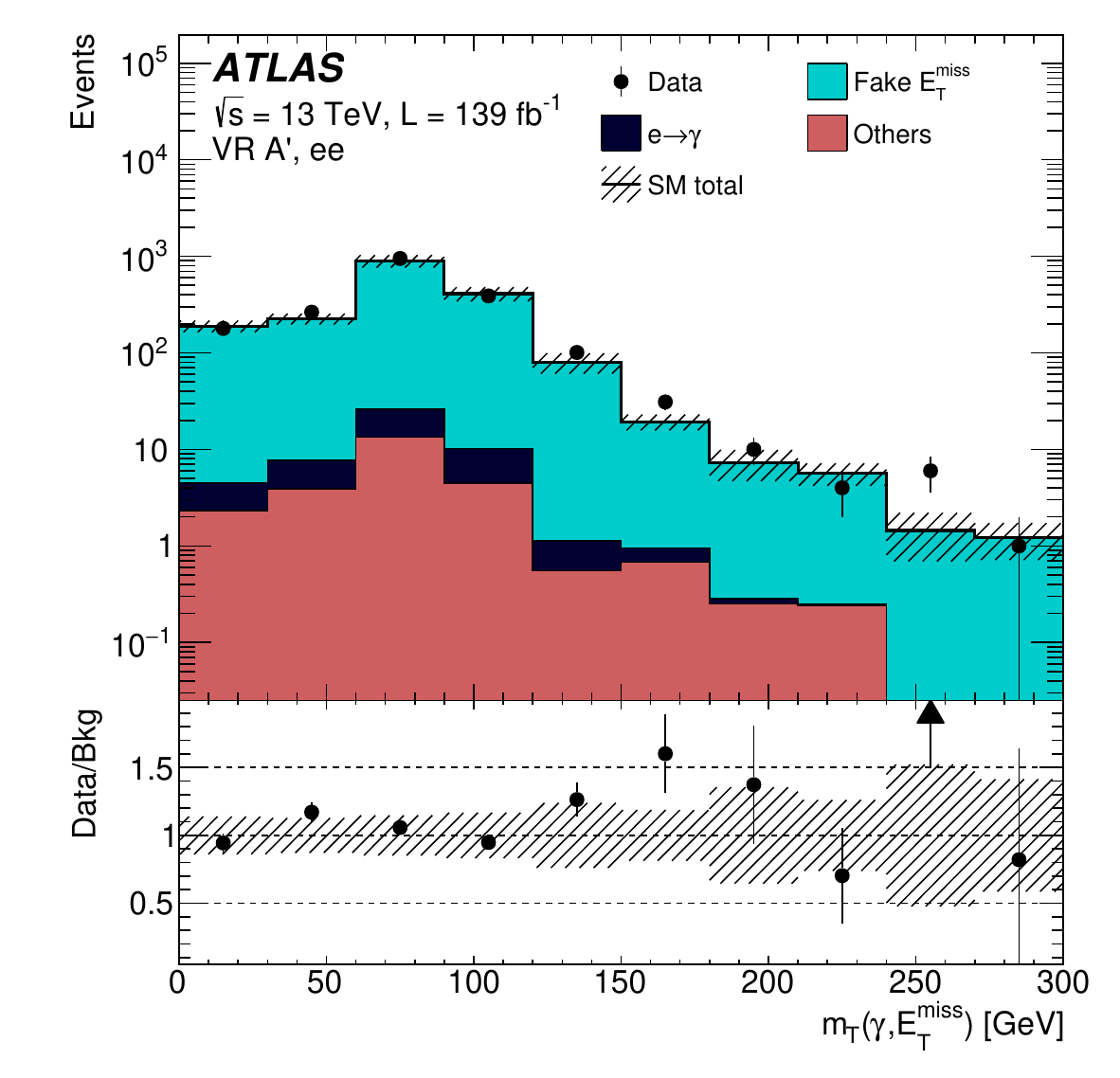}}
\subfloat[]{\includegraphics[scale=0.37]{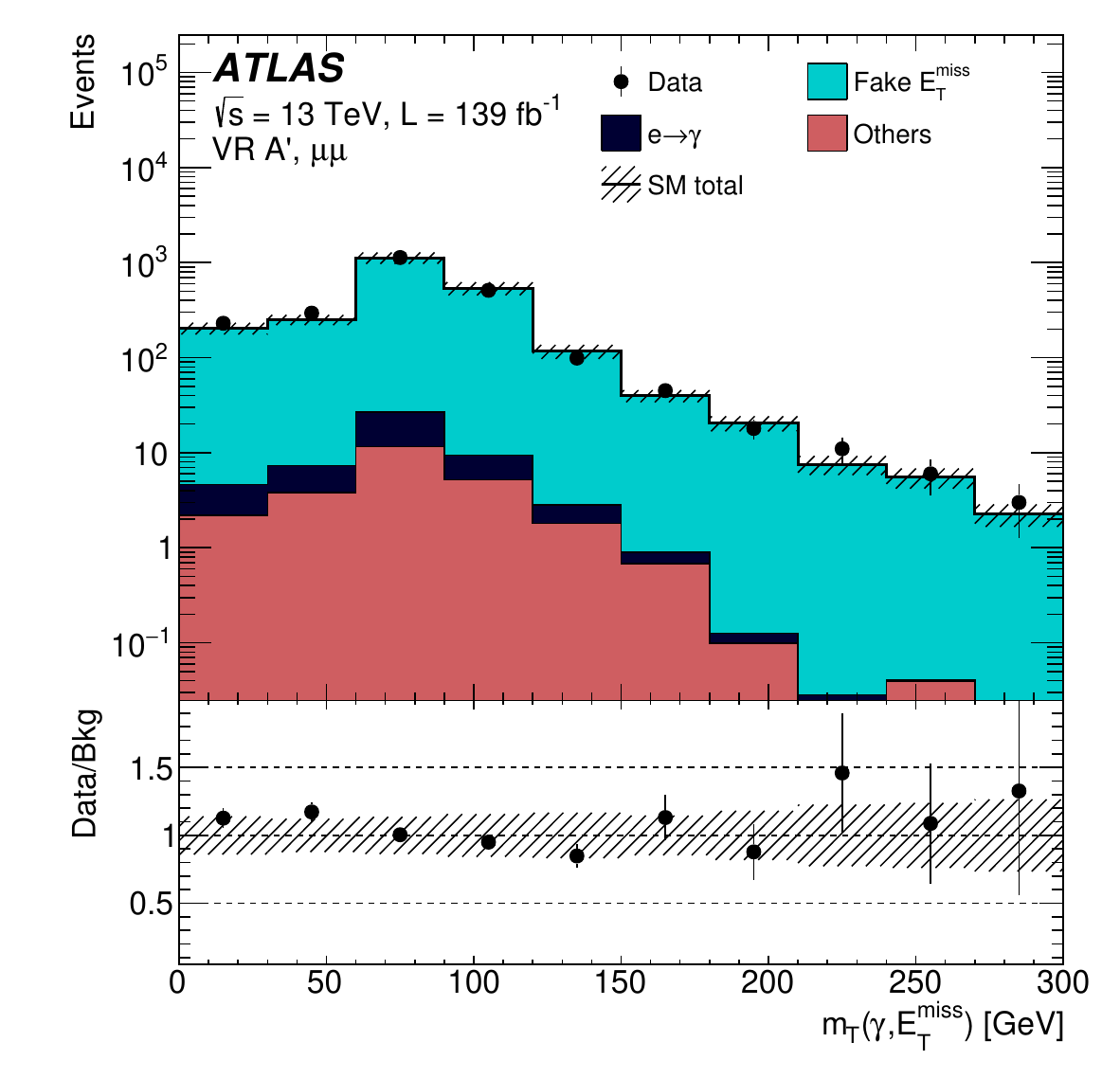}}
\caption{
Comparison between the expected background and data in the validation region $A'$, as a function of \mt,
for the $ee$ (a) and $\mu\mu$ (b) channels. The background yields from ``fake \met'' and ``$e\rightsquigarrow\gamma$'' are estimated with data-driven techniques. The other backgrounds are obtained from simulation and have been merged together. Uncertainties shown are statistical, both for data and for simulated backgrounds, while for the data-driven backgrounds the systematic uncertainties related to the method are also included.
}
\label{fig:bkg-Aprime-mT}
\end{figure}

As an overall check of the background estimates, the data-driven techniques described in Section~\ref{sec:bkg_fakemet} and Section~\ref{sec:bkg_elefake}
are applied in the VR $A'$ (Figure~\ref{fig:bkg_fakemet_ABCD}). The comparison between the expected background and data is displayed in Figures~\ref{fig:bkg-Aprime-ETmiss} and \ref{fig:bkg-Aprime-mT}, showing good agreement.
 
\section{Systematic uncertainties}
\label{sec:syst}

This section presents the various sources of systematic uncertainties affecting all levels of the analysis. They are categorised according to their origin, effects and the way they have been estimated.
 
\subsection{Experimental systematic uncertainties}
\label{sec:exp_sys}
The analysis is impacted by several uncertainties related to the detector resolution, inefficiencies and mis-measurements. They are grouped into the following categories: uncertainties in the luminosity, uncertainties in the trigger efficiencies and uncertainties related to the reconstruction of physics objects such as electrons, muons, jets, and \MET{}. The estimations of these uncertainties require a particular treatment as they affect the search via potential biases to the modelling of the
signal and background processes in the MC simulation. These systematic uncertainties can affect both the yield and the shape of the final observable (BDT classifier response) and are included in the statistical evaluation.
 
The uncertainty in the luminosity is 1.7\%~\cite{ATLAS-CONF-2019-021} and impacts the simulated yields of both the signal and backgrounds. This uncertainty was obtained using the LUCID-2 detector~\cite{LUCID2} for the primary luminosity measurements. In addition, systematic uncertainties resulting from effects of pile-up modeling, due to the reweighting procedure, are also considered.
 
Systematic uncertainties are estimated for lepton reconstruction and isolation efficiencies \cite{EGAM-2018-01,MUON-2018-03} and for the energy scale and resolution \cite{EGAM-2018-01}. Additional uncertainties on lepton trigger efficiencies are also considered to take into account differences between data and simulation \cite{TRIG-2018-01,TRIG-2018-05}.
 
For the photons, uncertainties in the reconstruction, isolation, energy scale and resolution are considered \cite{EGAM-2018-01,PERF-2017-02}.
 
For jets, uncertainties related to the energy and resolution \cite{JETM-2018-05,} as well as pile-up jet tagging \cite{ATL-PHYS-PUB-2014-001} are all taken into account. Specific systematic uncertainties are considered for heavy flavour tagging to correct for identification efficiencies of bottom, charm and light tagged jets \cite{ATL-PHYS-PUB-2017-013,ATL-PHYS-PUB-2020-014}.
 
For \MET{}, uncertainties associated with reconstructed objects are propagated and included in its calculation. Additional uncertainties associated to the \MET{} soft term  scale and resolution are independently evaluated \cite{PERF-2016-07}.
 
An additional uncertainty on the shape of the BDT classifier response for the fake \MET{}-induced background (Section~\ref{sec:bkg}) is considered. It is meant to account for inaccuracies in the shape of the $Z+$jets MC events characterized by high bin-to-bin fluctuations. To reduce these fluctuations a Gaussian smoothing algorithm is applied to the $Z+$jets distribution and the uncertainty is obtained from the difference between this varied shape and the one corresponding to the nominal $Z+$jets MC distribution together with the other MC simulated processes entering the fake \MET{} category.
 
Finally, the comparison between data and MC predictions in the ``top-VR'' shows a discrepancy at the level of $20\%$ for top-induced background (Section~\ref{sec:bkg_top_irred}). For this reason, in the treatment of such a background, a relative systematic uncertainty of $\pm20\%$ is considered.
 
\subsection{Theoretical systematic uncertainties}
\label{sec:theo_sys}
Theoretical uncertainties affect all simulated signal and background processes. They originate from the limited order in $\alpha_s$ or $\alpha_{EWK}$  at which the matrix elements are calculated, the matching of those calculations to parton showers and the uncertainty of the proton PDFs. The uncertainties from the variation of QCD scale and the variation of PDFs are considered for the signal and simulated background processes. Uncertainties for backgrounds entering different control (CR) and validation (VR) regions were also estimated.
 
For the $H \to \gamma\gamma_d$ signal, the total $ZH$ cross-section is calculated  up to NLO precision and is provided by the LHC Higgs Cross-Section Working Group \cite{deFlorian:2016spz}. NLO electroweak corrections reduce the cross section by 5.3\%.
 
This correction has been propagated to the BDT classifier response. The corresponding systematic uncertainties are extracted from the maximum and minimum variations of the NLO corrections and amount to 0.2-0.3\%.
The parton shower (PS) model effects on signal processes were  estimated through a comparison between \PYTHIA{}8 and \HERWIG{}7 showering models. The difference between the two algorithms in each BDT classifier response bin is taken as a systematic uncertainty on the PS model and amounts to 0.4-6.0\%  (0.2-36\%) for $qq\to ZH$ ($gg\to ZH$).
 
For the QCD renormalisation $\mu_R$ and factorisation $\mu_F$ scales a seven-point scale variation is considered, which amounts to varying the renormalisation and factorisation scales independently by factor of 1/2 and 2 around $\mu$ to the combinations of ($\mu_R$, $\mu_F$ ) = ($\mu$/2, $\mu$/2), (2$\mu$, 2$\mu$), ($\mu$, 2$\mu$), (2$\mu$, $\mu$), ($\mu$, $\mu$/2) and ($\mu$/2,$\mu$). The effect of these variations was propagated to the BDT classifier distributions. An envelope covering differences with respect to the nominal values is taken as the systematic uncertainty.

Depending on the $\gamma_d$ mass and the BDT bin, the uncertainties from PS vary from [-4,+5]\% ($qqZH$) to [-19,+25]\% ($ggZH$). Finally, uncertainties related to the choice of the PDF are computed by considering the yield predictions with full ensemble of 100 PDFs within the NNPDF set. The standard deviation of this set of yields is taken as the corresponding PDF uncertainty. The uncertainties related to the $\alpha_s$ variations are computed by considering the difference in the yield from two $\alpha_s$ variations (0.1195 and 0.1165). The $\alpha_s$ and PDF uncertainties are quadratically added and their impact varies from $\pm 0.4\%$ to $\pm 1.1\%$.
 
Similar approaches have also been applied to estimate theoretical uncertainties for background processes that were evaluated using MC predictions. Up and down variations with respect to nominal values for QCD ($\mu_R, \mu_F$) scales, PDF+$\alpha_s$ and PS algorithms were considered as systematic uncertainties for $VH,t\bar{t}H (H\to Z\gamma), t\bar{t}, t\bar{t}V, Wt\gamma$, single top-quark, $W\gamma^{QCD, EWK}$ and $VV\gamma$. Uncertainties from PDF+$\alpha_s$ amount to a maximum up to $\pm 10\%$ ($W\gamma$, single top-quark), QCD scale up to $\pm 30\%$ ($t\bar{t}V$) and PS up to $\pm 40\%$ ($W\gamma, t\bar{t}$). Finally, theoretical uncertainties associated with $Z\gamma$ and $Z+$jets were also evaluated as contributions from these processes need to be known in data-driven background  estimation.


\section{Results and Interpretation}
\label{sec:results}

 
To estimate the compatibility of the data with the SM expectations, as well as to extract upper limits at 95\% confidence level (CL) on the branching ratio of
$H\to\gamma\gamma_d$ a binned maximum likelihood fit is performed in the SR to the distribution of the BDT classifier response merging the $ee$ and $\mu\mu$ channels to obtain the best sensitivity.
 
The chosen binning in the SR is optimised to obtain the best expected sensitivity to the signal model, while also keeping low statistical uncertainties in each bin.
 
The likelihood function is built as a product of Poisson probability functions based on the expected signal and background yields in each BDT bin of the SR and in the single-bin $VV\gamma$ CR.  Two free parameters are included in the simultaneous likelihood fit: the branching ratio BR($H\to\gamma\gamma_d)$, which consists in the parameter of interest (POI) and multiplies the signal yield and the floating normalization factor for the $VV\gamma$ irreducible background $k_{VV\gamma}$, which is constrained by the $VV\gamma$ CR.  The systematic uncertainties are included as nuisance parameters, which are constrained by Gaussian distributions centered at zero with width equal to the corresponding uncertainty.
 
The dominant backgrounds, fake \met and electron faking photons are included in the fit taking the normalization from data-driven estimates described in Section \ref{sec:bkg}. To model the BDT shape in the fit, simulated events are used for fake \met, while for the background with an electron faking a photon data from the probe-electron CR (${e_p}$-CR) is considered after re-scaling each event by the appropriate fake rate $\Fey$.
 
The systematic uncertainties from data-driven methods are included in the fit as correlated among different BDT bins, as well as other experimental and theoretical systematic variations.
 
Two different fits are performed. The first corresponds to the background-only fit, in which the background predictions are determined in a fit to data assuming the presence of no signal. The second configuration instead allows for the presence of a specific signal and it is referred to as the model-dependent fit.
 
\begin{figure}
\centering
\subfloat[]{\includegraphics[scale=0.37]{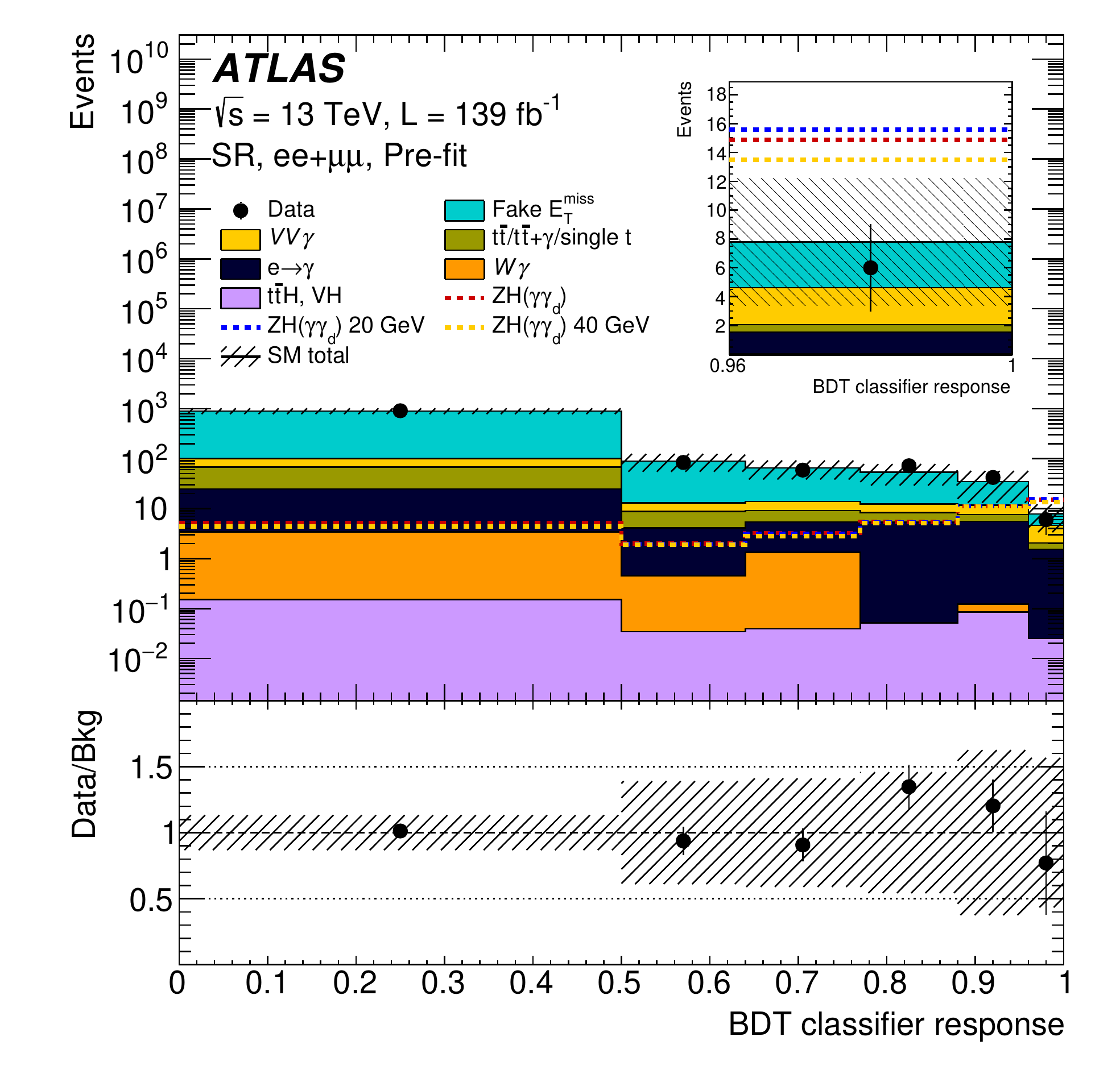}}
\subfloat[]{\includegraphics[scale=0.37]{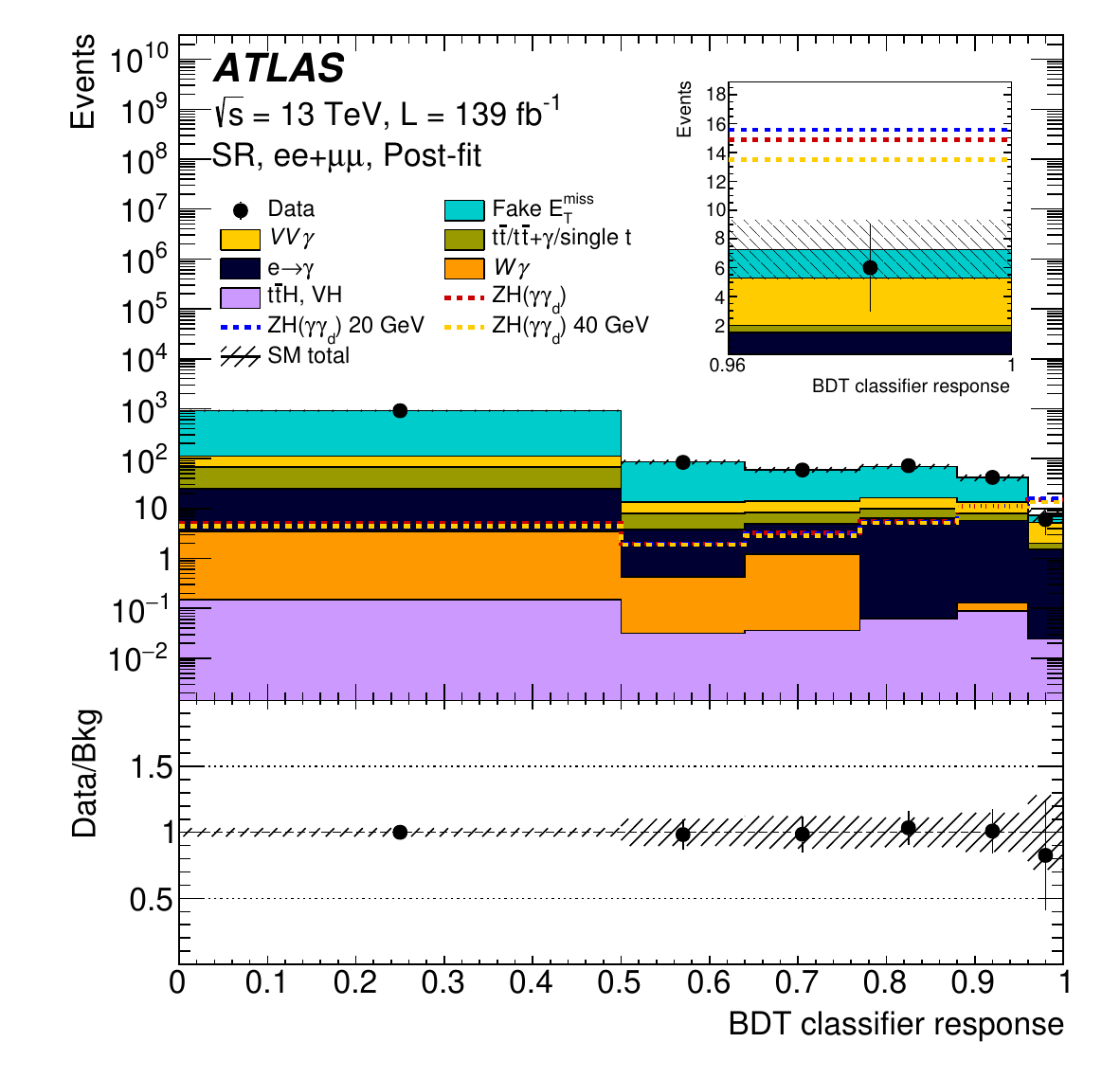}}
\caption{Distribution of the BDT classifier response in data and for the expected SM background before (a) and after (b) the background-only fit. The expectations for $ZH$, $H \to \gamma\gamma_d$ are also shown for the massless dark photon (red dashed line) and for dark photon mass values of 20~GeV (blue dashed line) and 40~GeV (yellow dashed line), assuming BR$(H\to \gamma\gamma_d)$ = 5\%.  A zoomed view of the last BDT bin with linear y-axis scale is also shown. Uncertainties shown are statistical for data, while for backgrounds include statistical and systematic sources determined by the multiple-bin fit. The lower panel shows the ratio of data to expected background event yields.}
\label{fig:BDT_fit}
\end{figure}

\begin{figure}
\centering
\includegraphics[scale=0.37]{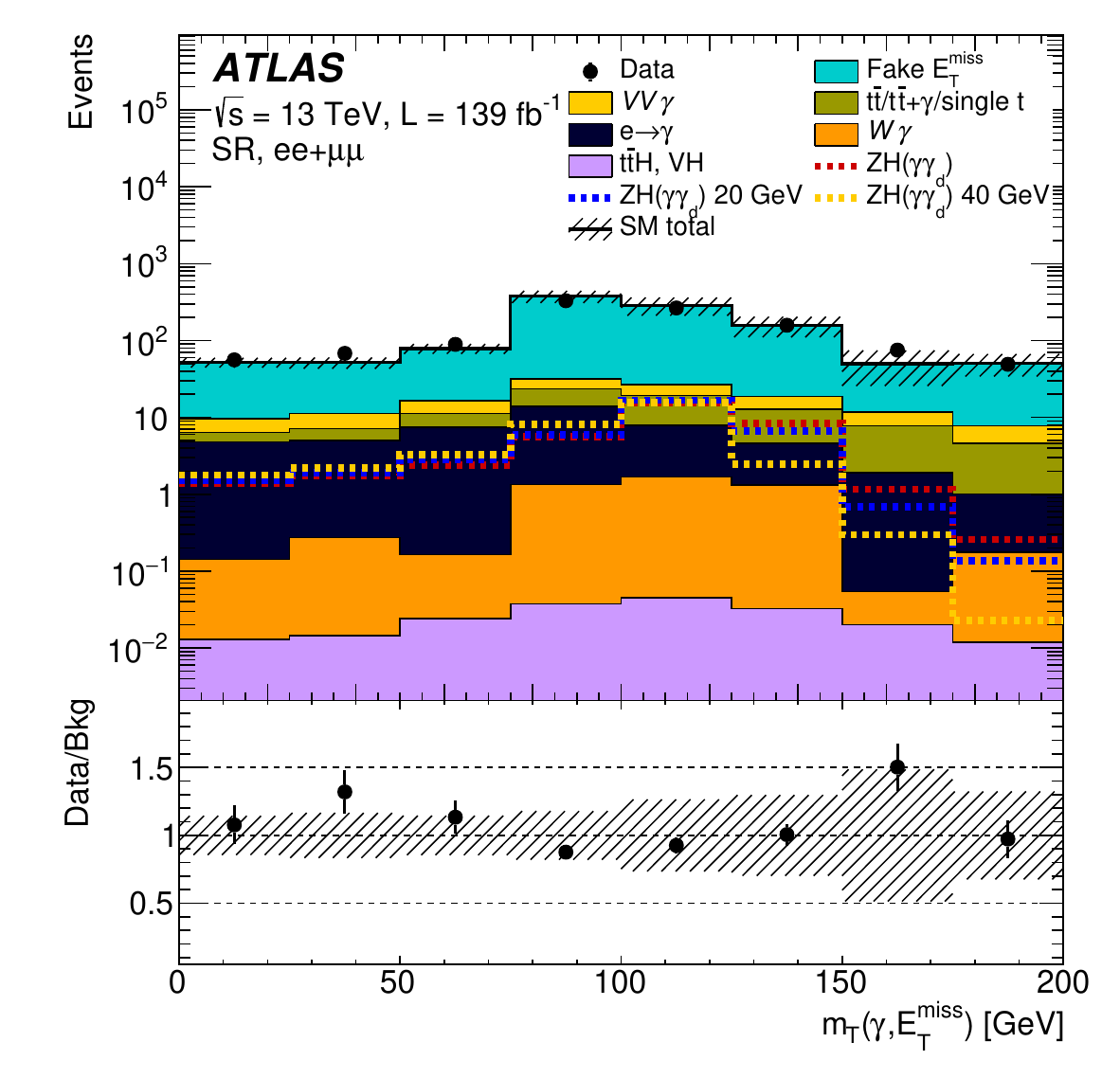}\\
\caption{Distribution of $m_\textnormal{T}$ in data and for the expected SM background before the background-only fit. The expectations for $ZH$, $H \to \gamma\gamma_d$ are also shown for the massless dark photon (red dashed line) and for dark photon mass values of 20~GeV (blue dashed line) and 40~GeV  (yellow dashed line), assuming BR$(H\to \gamma\gamma_d)$ = 5\%. Uncertainties shown are statistical for data, while for backgrounds include statistical and systematic sources determined by the multiple-bin fit. The lower panel shows the ratio of data to expected background event yields.}
\label{fig:mT_prefit}
\end{figure}

The background-only fit has been validated first in VR $A'$ to confirm the goodness of the SM expectations and the fitting procedure. The results of the background-only fit in SR are shown in Table \ref{tab:bkg_only_comb}, where observed and expected event yields are shown for all of the background processes considered in this analysis. The normalisation $k_{VV\gamma}$ factor is found to be 1.35 $\pm$ 0.38. The pre-fit and post-fit distributions of the BDT classifier response are shown in Figure  \ref{fig:BDT_fit}. Post-fit uncertainties are reduced thanks to the constraints on nuisance parameters, in particular the ones related to the fake \MET-induced background. Pre-fit distribution of $m_T$ is also shown in Figure  \ref{fig:mT_prefit}, being one of the most discriminant variables entering the BDT classifier.
 
The relative impact of each source of systematic uncertainty on the SM background estimates is summarised in Table \ref{tab:syst}. The purely statistical uncertainty of simulated samples is dominant, except in the last BDT bin, varying from about 3\% to 16\%. The largest systematic uncertainties in the last BDT bin are related to the shape of fake \met and to the jet energy scale and resolution corresponding respectively to 18\% and 13\%; uncertainty due to energy scale and resolution of electrons and photons corresponds to 5.6\%, while the same uncertainty for muons corresponds to 4.1\%. The other experimental and theoretical systematic uncertainties have a relative impact below about 3.5\% in all BDT bins.
 
\begin{table}
\caption{Observed event yields in 139~fb$^{-1}$ of data compared to expected yields from SM backgrounds obtained from the background-only fit for the $ee+\mu\mu$ channel in the SR and in the $VV\gamma$ CR. The total expected yields before the fit are also shown. The expected yields for the massless $\gamma_d$ signal are shown assuming BR$(H\to \gamma\gamma_d)$ = 5\%. The uncertainty includes both the statistical and systematic sources. The individual uncertainties can be correlated and do not necessarily add in quadrature to equal the total background uncertainty.}
\centering
\resizebox{\columnwidth}{!}{
\begin{tabular}{cccccccc}
\toprule
\toprule
BDT bin                      & SR 0 - 0.50            &  SR 0.50 - 0.64            &SR 0.64 - 0.77            &SR 0.77 -  0.88       &SR 0.88 - 0.96            &SR  0.96 - 1          &  $VV\gamma$ CR    \\
\midrule
Observed                       & $910$              & $84$              & $59$              & $72$              & $42$              & $6$           & $32$            \\
\midrule
Post-fit SM  background          & $910 \pm 29$   & $85.5 \pm 8.7$   & $59.9 \pm 7.3$   & $69.7 \pm 7.8$    & $41.6 \pm 6.1$    & $7.3 \pm 2.0$    & $31.4 \pm 5.4$       \\
\midrule
Fake $E_{T}^{miss}$  & $800 \pm 34$   & $72.1 \pm 8.3$   & $45.7 \pm 6.5$   & $53.2 \pm 7.1$   & $27.9 \pm 6.1$   & $2.0 \pm 1.9$ & $2.1_{-2.1}^{+3.5}$      \\
$e\rightsquigarrow\gamma$     & $21.5 \pm 2.0$  & $3.33 \pm 0.62$   & $3.75 \pm 0.74$   & $6.4 \pm 1.1$  & $5.7 \pm 1.4$   & $1.47 \pm 0.25$        & $1.24 \pm 0.07$       \\
$VV\gamma$                     & $44 \pm 12$          & $5.3 \pm 1.6$          & $5.8 \pm 1.7$          & $6.4 \pm 1.8$          & $5.7 \pm 1.9$          & $3.30 \pm 0.97$       & $27.3 \pm 6.4$           \\
$\ttbar$, $\ttbar\gamma$, single $t$                          & $42 \pm 15$          & $4.3 \pm 1.5$          & $3.4 \pm 1.2$          & $3.6 \pm 1.2$          & $2.13 \pm 0.80$          & $0.50 \pm 0.18$        & $0.63 \pm 0.22$            \\
$W\gamma$              & $3.3 \pm 1.5$          & $0.39 \pm 0.18$          & $1.18 \pm 0.55$          & $- $          & $0.04 \pm 0.02$          & $- $        & $-$       \\
$\ttbar H,~VH$                  & $0.15 \pm 0.02$          & $0.03 \pm 0.01$          & $0.04 \pm 0.01$          & $0.06 \pm 0.01$          & $0.09 \pm 0.03$          & $0.02 \pm 0.01$         & $0.17_{-0.17}^{+0.18}$     \\
\midrule
Pre-fit SM background                   & $900 \pm 120$          & $90 \pm 35$          & $65 \pm 27$          & $53 \pm 24$          & $35 \pm 22$          & $7.8 \pm 4.4$         & $24 \pm 4.7$      \\
 
\midrule
Signal ($ZH\to\gamma\gamma_d) $          & $5.1 \pm 1.3$          & $1.98 \pm 0.51$          & $3.2 \pm 1.0$          & $5.5 \pm 1.6$          & $11.1 \pm 3.1$          & $14.9 \pm 1.9$     &  $-$        \\

\bottomrule
\bottomrule
\end{tabular}
}
\label{tab:bkg_only_comb}
\end{table}
 
\begin{table}[!hbtp]
\centering
\small
\caption{Summary of the relative uncertainties in the background estimate for the BDT bins after the background-only fit. The individual uncertainties can be correlated and do not necessarily add in quadrature to equal the total background uncertainty.}
\label{tab:syst}
\begin{tabular}{lllllll}
\toprule
\toprule
\textbf{}      BDT  bin                & 0 - 0.50          & 0.50 - 0.64     &  0.64 - 0.77    &  0.77- 0.88    &   0.88 - 0.96     &  0.96 - 1               \\
\hline
\hline & [\%] & [\%] & [\%] & [\%] & [\%] & [\%] \\
Total (statistical+systematic) uncertainty         & $ 3.1$          & $ 10$          & $ 12$          & $ 11$          & $ 15$          & $28$
\\
Statistical uncertainty  & $ 3.1$          & $ 9.9$          & $ 12$          & $ 11$          & $ 14$          & $16$              \\
\hline
Fake \met shape & $0.17$  & $0.97$  & $0.40$  & $0.55$  & $2.8$  & $18$ \\
Jet $E$ scale and resolution & $0.02$  & $3.3$  & $2.1$  & $0.47$  & $2.1$  & $13$  \\
Electron, photon $E$ scale and resolution & $0.04$  & $0.45$  & $0.75$  & $0.46$  & $1.7$  & $5.6$ \\
Muon $E$ scale and resolution & $0.08$  & $0.17$  & $0.15$  & $0.91$  & $1.2$  & $4.1$  \\
Fake \met data-driven & $0.50$  & $0.28$  & $0.18$  & $0.04$  & $0.40$  & $3.5$   \\
\met soft term scale and resolution& $0.26$  & $0.16$  & $0.59$  & $0.49$  & $0.20$  & $2.8$  \\
Electron trigger/ID/iso/reco eff. & $0.01$  & $0.10$  & $0.10$  & $0.01$  & $0.17$  & $1.0$  \\
Muon trigger/ID/iso/reco eff. & $0.01$  & $0.07$  & $0.08$  & $0.06$  & $0.03$  & $0.84$  \\
Flavour tagging eff.& $<$  $0.01$  & $0.08$  & $0.10$  & $0.04$  & $0.02$  & $0.82$ \\
Electrons faking photons data-driven & $0.02$  & $0.08$  & $0.06$  & $0.06$  & $0.07$  & $0.73$  \\
Photon ID/iso/reco eff.& $0.01$  & $0.07$  & $0.08$  & $0.04$  & $0.09$  & $0.61$ \\
Reweighting of $\langle \mu \rangle$ in MC simulation      & $0.08$  & $0.10$  & $0.32$  & $0.46$  & $0.09$  & $0.48$ \\
Top normalization & $0.08$  & $0.06$  & $0.06$  & $0.02$  & $0.09$  & $0.13$  \\
\hline
Theoretical $VV\gamma$ & $0.04$  & $0.02$  & $0.16$  & $0.04$  & $0.13$  & $0.49$   \\
Theoretical fake \met & $0.05$  & $0.11$  & $0.12$  & $0.22$  & $0.29$  & $0.45$  \\
Theoretical top   & $0.09$  & $0.05$  & $0.17$  & $0.10$  & $0.04$  & $0.28$  \\
Theoretical $W\gamma$  & $0.04$  & $0.10$  & $0.18$  & $0.05$  & $0.13$  & $0.24$ \\
Theoretical Higgs   & $0.01$  & $0.05$  & $0.04$  & $0.02$  & $0.08$  & $0.05$   \\
\bottomrule
\bottomrule
\end{tabular}
\end{table}
 
\begin{figure}[h]
\centering
\includegraphics[width=0.8\textwidth]{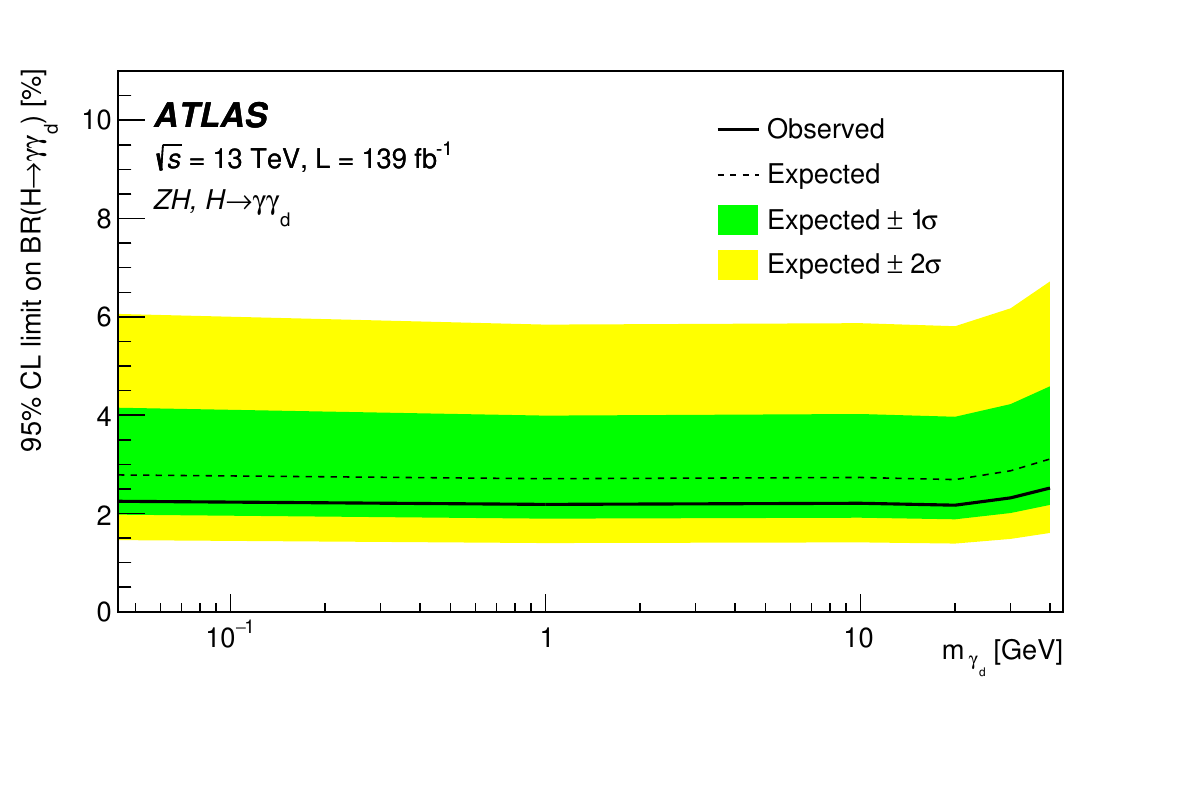}
\caption{Observed and expected exclusion limits at 95\% CL on BR($H\rightarrow \gamma\gamma_d$)  as function of the $\gamma_d$ mass. The green and yellow bands show respectively the $\pm 1\sigma$ and $\pm 2\sigma$ uncertainties.}
\label{fig:contour1VVyCR}
\end{figure}

The event yields in data are consistent with the predicted SM background event yields, as shown in Table \ref{tab:bkg_only_comb}. The model-dependent fit is therefore performed in order to extract upper limits at 95\%~CL on the branching ratio of the sought decay mode of the Higgs boson. These limits are based on the profile-likelihood-ratio test statistic~\cite{HistFitter} and CL$_\textrm{s}$ prescriptions~\cite{0954-3899-28-10-313}, evaluated using the asymptotic approximation~\cite{Cowan:2010js}. The fit is performed including the signal component of the Higgs boson production in $ZH$ with subsequent Higgs boson decays into $\gamma$ and $\gamma_d$. The results are provided for the massless dark photon, as well as for low dark photon mass values up to 40 GeV, as shown in Figure \ref{fig:contour1VVyCR}. The corresponding values are also reported in Table~\ref{tab:br_limits}. The observed (expected) upper limits on BR(H$\to \gamma\gamma_d)$  is at the level of 2.3\% (2.8\%), for massless $\gamma_d$  and varies slightly until mass values of 20~\GeV. The mass dependence of the limits become stronger beyond that value and the observed (expected) upper limit increases to about  2.5\% (3.1\%) at 40~\GeV.
This slight increase originates from the decreasing signal acceptance due the limited available phase space in this mass region.
 
\begin{table}[!hbtp]
\renewcommand{\arraystretch}{1.3}
\centering
\caption{Observed and expected  limits at 95\% CL on BR($H\to \gamma\gamma_d)$  for different values of the $\gamma_d$ mass for the $ee+\mu\mu$ channel.
The asymmetric error corresponds to the $\pm 1\sigma$}
\label{tab:br_limits}
\begin{tabular}{lcc}
\toprule
\toprule
$m_{\gamma_d}$                         &  BR($H\to \gamma\gamma_d)_{\mathrm{obs}}^{95\% \textnormal{ CL}}$ & BR($H\to \gamma\gamma_d)_{\mathrm{exp}}^{95\% \textnormal{ CL}}$  \\
\hline[GeV]    & [\%] & [\%]  \\
0&$2.28 $&$ 2.82^{+1.33}_{-0.84}$ \\
1&$2.19 $&$ 2.71^{+1.28}_{-0.81}$ \\
10&$2.21 $&$ 2.73^{+1.31}_{-0.82}$ \\
20&$ 2.17 $&$ 2.69^{+1.29}_{-0.81}$ \\
30&$ 2.32 $&$ 2.87^{+1.36}_{-0.86}$ \\
40&$ 2.52 $&$ 3.11^{+1.48}_{-0.93}$ \\
\bottomrule
\bottomrule
\end{tabular}
\label{tab:limit}
\end{table}


\section{Conclusion}
\label{sec:conclusion}

 
A search for dark photon candidates arising from semi-visible Standard Model Higgs boson decay $H\to \gamma\gamma_d$ is performed. The Higgs boson production in association with a $Z(\to\ell^+\ell^-)$ boson is exploited, which benefits from a relatively clean signal and high-efficiency lepton triggers. The search uses $pp$ collision data collected by the ATLAS experiment at a centre-of-mass energy of $\sqrt{s} = 13$~\TeV{} between 2015 and 2018 and corresponding to an integrated luminosity of 139~fb$^{-1}$. Data-driven techniques are optimized to estimate the main backgrounds from processes characterized by fake \MET{} and electrons misidentified as photons, while the normalization of the irreducible background is obtained using MC simulations constrained by data in a dedicated control region.
 
The sensitivity of the search is enhanced thanks to a Boosted Decision Tree algorithm that permits the construction of  the discriminant kinematic observable. No excess of events above the SM expectation is found.  Therefore, limits on the branching ratio of a SM Higgs boson decaying to a photon and a dark photon can be set. For massless $\gamma_d$, an observed (expected) upper limit on BR($H\to \gamma\gamma_d$) of 2.28\% (2.82$^{+1.33}_{-0.84}$\%) is set at 95\%~CL. For massive $\gamma_d$, the observed (expected) upper limits are found to be within the [2.19,2.52]\% ([2.71,3.11]\%) range for masses spanning from 1~\GeV~ to 40~\GeV.


\section*{Acknowledgements}


We thank CERN for the very successful operation of the LHC, as well as the
support staff from our institutions without whom ATLAS could not be
operated efficiently.
 
We acknowledge the support of
ANPCyT, Argentina;
YerPhI, Armenia;
ARC, Australia;
BMWFW and FWF, Austria;
ANAS, Azerbaijan;
CNPq and FAPESP, Brazil;
NSERC, NRC and CFI, Canada;
CERN;
ANID, Chile;
CAS, MOST and NSFC, China;
Minciencias, Colombia;
MEYS CR, Czech Republic;
DNRF and DNSRC, Denmark;
IN2P3-CNRS and CEA-DRF/IRFU, France;
SRNSFG, Georgia;
BMBF, HGF and MPG, Germany;
GSRI, Greece;
RGC and Hong Kong SAR, China;
ISF and Benoziyo Center, Israel;
INFN, Italy;
MEXT and JSPS, Japan;
CNRST, Morocco;
NWO, Netherlands;
RCN, Norway;
MEiN, Poland;
FCT, Portugal;
MNE/IFA, Romania;
MESTD, Serbia;
MSSR, Slovakia;
ARRS and MIZ\v{S}, Slovenia;
DSI/NRF, South Africa;
MICINN, Spain;
SRC and Wallenberg Foundation, Sweden;
SERI, SNSF and Cantons of Bern and Geneva, Switzerland;
MOST, Taiwan;
TENMAK, T\"urkiye;
STFC, United Kingdom;
DOE and NSF, United States of America.
In addition, individual groups and members have received support from
BCKDF, CANARIE, Compute Canada and CRC, Canada;
PRIMUS 21/SCI/017 and UNCE SCI/013, Czech Republic;
COST, ERC, ERDF, Horizon 2020, ICSC-NextGenerationEU and Marie Sk{\l}odowska-Curie Actions, European Union;
Investissements d'Avenir Labex, Investissements d'Avenir Idex and ANR, France;
DFG and AvH Foundation, Germany;
Herakleitos, Thales and Aristeia programmes co-financed by EU-ESF and the Greek NSRF, Greece;
BSF-NSF and MINERVA, Israel;
Norwegian Financial Mechanism 2014-2021, Norway;
NCN and NAWA, Poland;
La Caixa Banking Foundation, CERCA Programme Generalitat de Catalunya and PROMETEO and GenT Programmes Generalitat Valenciana, Spain;
G\"{o}ran Gustafssons Stiftelse, Sweden;
The Royal Society and Leverhulme Trust, United Kingdom.
 
The crucial computing support from all WLCG partners is acknowledged gratefully, in particular from CERN, the ATLAS Tier-1 facilities at TRIUMF (Canada), NDGF (Denmark, Norway, Sweden), CC-IN2P3 (France), KIT/GridKA (Germany), INFN-CNAF (Italy), NL-T1 (Netherlands), PIC (Spain), ASGC (Taiwan), RAL (UK) and BNL (USA), the Tier-2 facilities worldwide and large non-WLCG resource providers. Major contributors of computing resources are listed in Ref.~\cite{ATL-SOFT-PUB-2023-001}.


\clearpage
\clearpage

\printbibliography
 
\clearpage
 
\begin{flushleft}
\hypersetup{urlcolor=black}
{\Large The ATLAS Collaboration}

\bigskip

\AtlasOrcid[0000-0002-6665-4934]{G.~Aad}$^\textrm{\scriptsize 102}$,
\AtlasOrcid[0000-0002-5888-2734]{B.~Abbott}$^\textrm{\scriptsize 120}$,
\AtlasOrcid[0000-0002-7248-3203]{D.C.~Abbott}$^\textrm{\scriptsize 103}$,
\AtlasOrcid[0000-0002-1002-1652]{K.~Abeling}$^\textrm{\scriptsize 55}$,
\AtlasOrcid[0000-0002-8496-9294]{S.H.~Abidi}$^\textrm{\scriptsize 29}$,
\AtlasOrcid[0000-0002-9987-2292]{A.~Aboulhorma}$^\textrm{\scriptsize 35e}$,
\AtlasOrcid[0000-0001-5329-6640]{H.~Abramowicz}$^\textrm{\scriptsize 151}$,
\AtlasOrcid[0000-0002-1599-2896]{H.~Abreu}$^\textrm{\scriptsize 150}$,
\AtlasOrcid[0000-0003-0403-3697]{Y.~Abulaiti}$^\textrm{\scriptsize 117}$,
\AtlasOrcid[0000-0003-0762-7204]{A.C.~Abusleme~Hoffman}$^\textrm{\scriptsize 137a}$
% ...
\bigskip
\\

% ...
$^{29}$Physics Department, Brookhaven National Laboratory, Upton NY; United States of America.\\
% ...
$^{35}$$^{(a)}$Facult\'e des Sciences Ain Chock, R\'eseau Universitaire de Physique des Hautes Energies - Universit\'e Hassan II, Casablanca;$^{(b)}$Facult\'{e} des Sciences, Universit\'{e} Ibn-Tofail, K\'{e}nitra;$^{(c)}$Facult\'e des Sciences Semlalia, Universit\'e Cadi Ayyad, LPHEA-Marrakech;$^{(d)}$LPMR, Facult\'e des Sciences, Universit\'e Mohamed Premier, Oujda;$^{(e)}$Facult\'e des sciences, Universit\'e Mohammed V, Rabat;$^{(f)}$Institute of Applied Physics, Mohammed VI Polytechnic University, Ben Guerir; Morocco.\\
% ...
$^{55}$II. Physikalisches Institut, Georg-August-Universit\"{a}t G\"ottingen, G\"ottingen; Germany.\\
% ...
$^{102}$CPPM, Aix-Marseille Universit\'e, CNRS/IN2P3, Marseille; France.\\
$^{103}$Department of Physics, University of Massachusetts, Amherst MA; United States of America.\\
% ...
$^{117}$Department of Physics, New York University, New York NY; United States of America.\\
% ...
$^{120}$Homer L. Dodge Department of Physics and Astronomy, University of Oklahoma, Norman OK; United States of America.\\
% ...
$^{137}$$^{(a)}$Departamento de F\'isica, Pontificia Universidad Cat\'olica de Chile, Santiago;$^{(b)}$Millennium Institute for Subatomic physics at high energy frontier (SAPHIR), Santiago;$^{(c)}$Instituto de Investigaci\'on Multidisciplinario en Ciencia y Tecnolog\'ia, y Departamento de F\'isica, Universidad de La Serena;$^{(d)}$Universidad Andres Bello, Department of Physics, Santiago;$^{(e)}$Instituto de Alta Investigaci\'on, Universidad de Tarapac\'a, Arica;$^{(f)}$Departamento de F\'isica, Universidad T\'ecnica Federico Santa Mar\'ia, Valpara\'iso; Chile.\\
% ...
$^{150}$Department of Physics, Technion, Israel Institute of Technology, Haifa; Israel.\\
$^{151}$Raymond and Beverly Sackler School of Physics and Astronomy, Tel Aviv University, Tel Aviv; Israel.\\
% ...
$^{a}$ Also Affiliated with an institute covered by a cooperation agreement with CERN.\\
$^{b}$ Also at An-Najah National University, Nablus; Palestine.\\
$^{c}$ Also at Borough of Manhattan Community College, City University of New York, New York NY; United States of America.\\
$^{d}$ Also at Bruno Kessler Foundation, Trento; Italy.\\
$^{e}$ Also at Center for High Energy Physics, Peking University; China.\\
$^{f}$ Also at Center for Interdisciplinary Research and Innovation (CIRI-AUTH), Thessaloniki; Greece.\\
% ...

\end{flushleft}


\end{document}